%% file: main.tex
\documentclass{article}
\usepackage[utf8]{inputenc}
\usepackage[T1]{fontenc}
\usepackage{microtype}
\usepackage{graphicx}
\usepackage{booktabs}
\usepackage{amsmath}
\usepackage{amsfonts}
\usepackage{xcolor}
\usepackage{colortbl}
\usepackage{tabularx}
\usepackage{listings}
\usepackage{float}
\usepackage{xurl}
\usepackage{hyperref}
\usepackage[author]{preprint}
\makeatletter
\renewcommand{\pp@appearing}{Preprint.}
\makeatother
\pptitlerunning{Execution-Grounded Specialization for Token-Efficient Skill Use}
\begin{document}
\twocolumn[
\pptitle{How Much Skill Orchestration Can LLM Agents Compile Away? Execution-Grounded
Specialization for Token-Efficient Skill Use}
\begin{ppauthorlist}
\ppauthor{Jiayi Ni}{nv}
\ppauthor{Zac Wang}{nv}
\ppauthor{Alex Zhang}{nv}
\ppauthor{Zheng Liu}{nv}
\ppauthor{Roopa Prabhu}{nv}
\end{ppauthorlist}
\ppaffiliation{nv}{NVIDIA}
\ppkeywords{LLM agents, skill reuse, execution-grounded compilation, inference efficiency}
\vskip 0.3in
\begin{abstract}
Reusable skills guide large language model (LLM) agents through recurring requests, but each
invocation still requires the model to interpret the skill and reconstruct its orchestration.
Can agents automatically compile this orchestration into an executable fast path without human
intervention? We first study \emph{zero-shot skill specialization}, which compiles without execution evidence,
and find that despite the skill's procedural guidance, incorrect assumptions about runtime bindings
can make generated plans brittle and recovery costly. This motivates SkillCache, a
\emph{one-trace skill specialization} method that compiles a completed invocation into a
parameterized capsule of programs and natural-language steps, replays it before reuse, and drives it with a plan
executor that returns control to the model only when needed, keeping the source skill for recovery.
With the Claude Code harness, warm execution of the same request reduces tokens on 87 SkillsBench tasks by 81\%
with Claude Sonnet~4.6 and by 74\% with Claude Opus~4.8 at comparable
reward, and capsules stay effective under controlled request changes. Capsules also transfer across models,
raising three open-weight models' mean reward by 12--25 percentage points while reducing their
tokens by 72--81\%, suggesting that execution-grounded compilation is a promising approach to
efficient repeated skill use and cross-model transfer.
\end{abstract}
]
{\let\thefootnote\relax\footnotetext{\hspace*{-\footnotesep}\textsuperscript{1}NVIDIA.\ \\ Preprint.}}
\input{introduction}
\input{background}
\input{compilation_timing}
\input{skillcache_method}
\input{compiling_capsule}

\input{executing_capsule}
\input{experiments_results}
\input{limitations}
\input{related_work}
\input{conclusion}
\clearpage
\bibliography{references}
\bibliographystyle{preprint}
\input{ai_usage_disclosure}
\clearpage
\appendix
\input{appendix}
\input{appendix_prompts}
\end{document}

%% file: introduction.tex
\section{Introduction}
\label{sec:introduction}

Many large language model (LLM) agent workloads contain recurring requests of the same type, such as the same analysis over
new datasets~\citep{zhang2025apc,li2026preact,elyadouni2026tracecompiler}. Agent Skills package instructions and supporting files as
reusable procedures for such requests~\citep{anthropic2025equipping,anthropicAgentSkills}.
Well-written skills can improve overall task accuracy and sometimes reduce exploration and
tokens~\citep{li2026skillsbench,zhong2026skilllearnbench}, yet the agent still needs to plan and reason over the matched skill on every invocation. In particular, irrelevant skill context and
repeated reasoning add runtime overhead~\citep{xu2026skillsmith}.

Recent source-level compilers transform skill specifications into structured or executable forms
\citep{ouyang2026skcc,xu2026skillsmith,dantanarayana2026sigil}, but operate mainly on skill packages instead of specializing orchestration for a concrete request. Other methods evolve or optimize skills. Some use execution trajectories and task-level
rewards~\citep{mi2026skillpro,xia2026skillrl,yang2026skillmaster}, held-out rollout evaluation~\citep{yang2026skillopt},
or official verifier feedback~\citep{rao2026skillopt} to learn, select, or repair updates, yet such
signals may be unavailable in deployment. Human-authored skills also outperform the automatic alternatives evaluated by
\citet{zhong2026skilllearnbench}. Thus, we study a complementary setting that preserves the
human-authored skill as the source of truth and stores execution-derived specialization separately.

Building on systems that separate planning from tool execution
\citep{xu2023rewoo,kim2024llmcompiler,erdogan2025planandact}, we first study \emph{zero-shot skill
specialization}, where an agent compiles the skill and request into a reusable capsule of programs and natural-language steps before
execution. The goal is to pay the compilation cost once and reuse the resulting capsule instead of repeatedly
interpreting the skill. On 87 SkillsBench tasks~\citep{li2026skillsbench}, however, compiling and executing once uses
more tokens than direct skill use on 58 of the 84 tasks without a timeout, and $1.24\times$ as many per task in geometric mean. On the 32 tasks
with a compiled-program failure, execution alone uses $0.92\times$ the direct tokens, against
$0.71\times$ over all tasks. Including compilation raises the ratio on these 32 tasks to $1.34\times$.

\begin{figure}[t]
  \centering
  \includegraphics[width=\columnwidth]{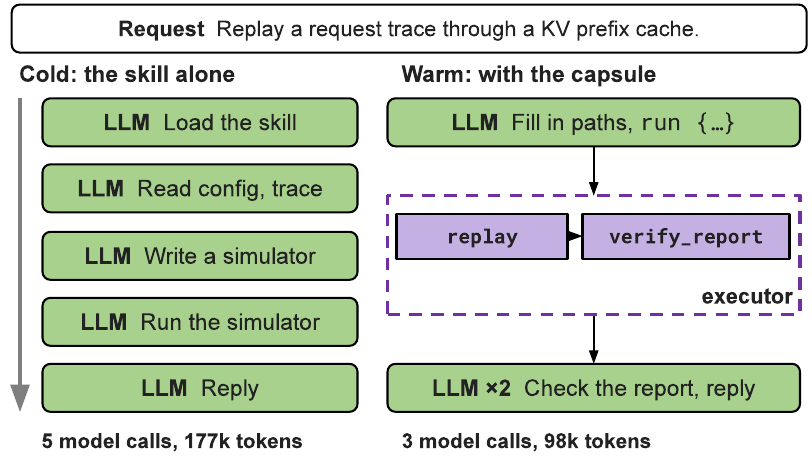}
  \caption{The same request, run cold and warm (SkillsBench \texttt{llm-prefix-cache-replay}, Claude
  Opus~4.8). In the warm run, the model fills in the paths from the request, and the executor runs
  the compiled programs.}
  \label{fig:cold-warm}
\end{figure}

These failures motivate using execution evidence alongside the skill source during specialization.
Prior work derives reusable procedures or executable artifacts from observed behavior
\citep{chen2026skillcraft,ni2026trace2skill,chen2026skilldroid,li2026preact,elyadouni2026tracecompiler}, sometimes by aggregating
multiple trajectories or using an external evaluator to confirm success. In our setting, the supplied skill already provides the general procedure. We
introduce \emph{SkillCache}, a \emph{one-trace skill specialization} method that uses a single
completed trajectory to recover the runtime bindings and execution structure. An evaluator-blind compiler agent automatically
compiles the unchanged skill and trace into a parameterized capsule of programs and natural-language
steps. At reuse, SkillCache drives the capsule with a plan executor that runs consecutive programs
and returns control to the model only for its input, its judgment, or a failed program, so the model
no longer orchestrates each call. The source skill remains available for recovery.
We call the initial execution \emph{cold} and subsequent capsule-augmented executions \emph{warm},
and Figure~\ref{fig:cold-warm} shows both on an example request. Our contributions are as follows:
\begin{itemize}
  \item We first show that specializing a skill from the request and skill alone does not reliably reduce
  tokens, because the compiler may guess runtime bindings incorrectly and failed programs are expensive to
  recover.
  \item We thus present SkillCache, a one-trace skill specialization method that turns one completed
  execution into a capsule of programs and natural-language steps and replays it before reuse for qualification.
  \item Building on the capsule's structure, we also present an executor-driven design, which runs
  the programs and passes values between them, returning control to the model only when needed.
  \item We evaluate SkillCache on 87 SkillsBench tasks. Averaged over three runs, warm execution
  reduces tokens by 81\% with Claude Sonnet~4.6 and by 74\% with Claude Opus~4.8 at comparable reward, repaying the dollar cost of compilation after 1.1 and 1.5 reuses in aggregate. Capsules also transfer to three open-weight
  models, raising their mean reward by 12--25 percentage points. We further show that capsules stay
  effective under controlled request changes and that the executor-driven design and replay both
  contribute to the savings.
\end{itemize}

%% file: background.tex
\section{Background}
\label{sec:background}

\subsection{Agent Skills and Their Orchestration Cost}
\label{sec:background-skills}

LLM agents combine reasoning with environment actions and external
tools~\citep{yao2023react,schick2023toolformer,patil2024gorilla}. Agent harnesses such as Claude
Code~\citep{anthropic2025claudecode}, Codex~\citep{openai2025codex}, and
OpenClaw~\citep{openclawDocs} manage context, tools, and multi-turn execution.
Agent Skills extend an agent with reusable procedures, each a directory with a
\texttt{SKILL.md} file of instructions and optional scripts and reference
files~\citep{anthropic2025equipping,anthropicAgentSkills}. The model first reads each
skill's name and description, decides whether the skill applies, and opens its instructions and
files when it needs them. It then interprets the skill and orchestrates its procedure through
tool calls in the harness.

Each model call takes the
instructions and the execution history so far as input, so input tokens accumulate across calls.
Provider prompt caching makes repeated prefixes cheaper~\citep{lumer2026dontbreak}, but their
cost still grows with every call. Each call also generates new output tokens and adds latency. In the cold executions of our main study (Section~\ref{sec:evaluation}), a request takes a median of 14 calls with both Sonnet~4.6 and Opus~4.8.

Many of these calls do mechanical work, such as composing a command or copying a value from one
output into the next. Only
judgment needs a model, yet every recurring request has the model redo the
mechanical work. Together, these observations suggest compiling the mechanical work of a skill invocation into programs that later requests reuse, and involving the model only where judgment is needed.

\subsection{Tracing JIT Compilation}
\label{sec:background-jit}

Tracing just-in-time (JIT) compilers reuse compiled code for frequently executed paths within a program
run~\citep{gal2009tracemonkey}. While interpreting a program, they record a path, called a trace, and only then generate
code specialized to the types observed along it. This specialization assumes that later executions
follow the same path with compatible types, and guards check these assumptions at runtime. If a guard
fails, execution leaves the compiled trace and can return to the interpreter through deoptimization.
The compiler must preserve the program's meaning across specialization and
deoptimization~\citep{fluckiger2018correctness}. In Section~\ref{sec:skillcache}, we detail how
SkillCache brings the idea of tracing JIT compilation to skill orchestration in LLM agents.

%% file: compilation_timing.tex
\section{Can Skills Be Specialized without Execution Evidence?}
\label{sec:timing}

\subsection{Zero-Shot Skill Specialization}

A skill-guided invocation comprises a request $q$, supplied skill packages $S$, and an initial workspace
$W_0$. Given execution evidence $h$, a compiler $C_{\theta_c}$ produces a capsule $z_h$,
and an executor $E_{\theta_e}$ uses it to interact with tools:
\begin{equation}
  \begin{gathered}
    z_h\sim C_{\theta_c}(q,S,h),\\
    (y_h,W_h,\tau_h)\sim E_{\theta_e}(q,S,W_0,z_h).
  \end{gathered}
  \label{eq:invocation}
\end{equation}
Here $\theta_c$ and $\theta_e$ denote the compiler and executor configurations, $y_h$ is the response, and
$\tau_h=((a_1,o_1),\ldots,(a_T,o_T))$ is the observable trajectory. Let $m_h$ and $k_h$ denote the provider-reported tokens of this
execution and of its compilation, and $m_0$ those of direct skill execution.

Zero-shot specialization compiles without observing task execution:
\begin{equation}
  h_{\mathrm{zero}}=\emptyset,\quad
  z_{\mathrm{zero}}\sim C_{\theta_c}(q,S,\emptyset).
  \label{eq:zero-shot}
\end{equation}
The compiler must decide which branches and bindings can be fixed ahead of time, and incorrect
assumptions can cause failures and recovery. This raises the question of whether a skill and request alone
support reliable specialization.

Since the compile call precedes zero-shot execution, the primary comparison includes compilation
and execution, $k_{\mathrm{zero}}+m_{\mathrm{zero}}$. If the capsule is persisted, its tokens over
$N$ compatible invocations are $k_{\mathrm{zero}}+Nm_{\mathrm{zero}}$, compared with $Nm_0$ for direct
skill execution, assuming fixed per-invocation costs. Persisting the capsule saves tokens overall when
$N(m_0-m_{\mathrm{zero}})>k_{\mathrm{zero}}$.

\subsection{Evaluating Zero-Shot Specialization}
\label{sec:preexecution-study}

We run 87 SkillsBench tasks~\citep{li2026skillsbench} with Sonnet~4.6 using the Claude Code
harness~\citep{anthropic2025claudecode}. The direct and zero-shot conditions use the same skills, model, and environment, and the zero-shot compiler sees only the
request and skills (Appendix~\ref{app:zero-shot}). We leave out
the three tasks where an execution timed out, count the tasks on which zero-shot uses fewer or more
tokens than direct execution, and apply a two-sided sign test.

\textbf{Results.}
Table~\ref{tab:preexecution} shows that, counting execution alone, zero-shot uses $0.71\times$ the
direct tokens per task. However, the compile call outweighs this
saving. Once it is included, zero-shot uses $1.24\times$ the direct tokens per task, and 58 of 84 tasks
use more. Total tokens decrease by 5\%, but one task holds 18\% of the
direct tokens, and without it they increase by 11\% (Appendix~\ref{app:zero-shot}). Reward rises on 10 tasks and decreases on 12 ($p=0.83$). A post-hoc reading of each trajectory separates two failure modes. On 10 tasks, a program returned a
wrong answer without an error, and 9 of these scored 0. On 32 others, a compiled program failed
observably and the model recovered through direct skill use. Execution alone then uses $0.92\times$
the direct tokens per task, and $1.34\times$ with the compile
call.

\begin{table}[tb]
  \caption{Zero-shot against direct execution on 84 tasks (Sonnet~4.6). \emph{Ratio} is the geometric
  mean of zero-shot over direct tokens, and \emph{Fewer/more} counts tasks. A star marks $p<0.005$ in a sign
  test. \emph{Failure} marks observable program failures.}
  \label{tab:preexecution}
  \centering
  \small
  \setlength{\tabcolsep}{4pt}
  \input{tables/zero_shot_table.tex}
\end{table}

Even when reuse amortizes the compile call, zero-shot savings remain uneven across tasks
(Section~\ref{sec:eval-ablations}).
This motivates \emph{one-trace skill specialization}, which uses the branch and interfaces observed
in a completed trajectory $\tau_{\mathrm{obs}}$ to specialize the unchanged source skill:
\begin{equation}
  z_{\mathrm{zero}}\sim C_{\theta_c}(q,S,\emptyset),\quad
  z_{\mathrm{trace}}\sim C_{\theta_c}(q,S,\tau_{\mathrm{obs}}).
  \label{eq:one-trace-specialization}
\end{equation}

\begin{figure*}[t]
  \centering
  \includegraphics[width=\linewidth]{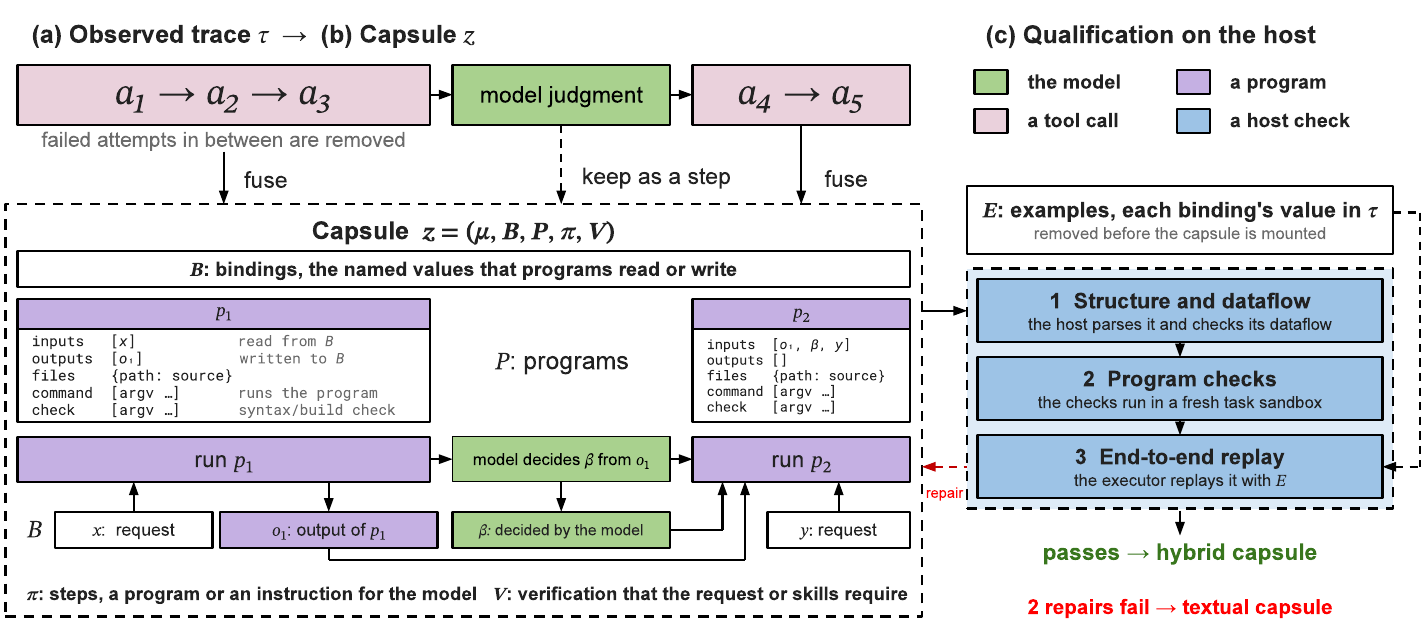}
  \caption{Compiling an execution into a capsule. (a) An observed execution. (b) The compiler fuses
  each run of tool calls into a program and keeps the judgment as a natural-language step. (c) The
  host admits the capsule only after three gates.}
  \label{fig:capsule-compilation}
\end{figure*}

%% file: tables/zero_shot_table.tex
\begin{tabular*}{\linewidth}{@{\extracolsep{\fill}}lrcccc@{}}
\toprule
 & & \multicolumn{2}{c}{Compile + execution} & \multicolumn{2}{c}{Execution only} \\
\cmidrule(lr){3-4}\cmidrule(l){5-6}
Tasks & $n$ & Ratio & Fewer/more & Ratio & Fewer/more \\
\midrule
All & 84 & $1.24\times$ & 26/58$^{*}$ & $0.71\times$ & 54/30 \\
No failure & 52 & $1.19\times$ & 17/35 & $0.60\times$ & 37/15$^{*}$ \\
Failure & 32 & $1.34\times$ & 9/23 & $0.92\times$ & 17/15 \\
\bottomrule
\end{tabular*}

%% file: skillcache_method.tex
\section{SkillCache Design Overview}
\label{sec:skillcache}

SkillCache performs branch-level specialization of a supplied skill using one completed invocation.
It works in three phases. \textbf{(1)~Observe}: It records the cold invocation's observable trace
$\tau_{\mathrm{obs}}$, which holds the tool calls with their arguments, results, and errors, and the
agent's messages. The trace identifies the branch the skill took and carries its runtime arguments,
intermediate observations, and recovery actions. \textbf{(2)~Compile}: An isolated compiler turns $\tau_{\mathrm{obs}}$ into a \emph{capsule}, fusing adjacent
deterministic tool calls into programs, keeping judgments as natural-language steps, and lifting
request-dependent values into bindings (Figure~\ref{fig:capsule-compilation}). The host admits a capsule only after its programs pass a replay in the task's
sandbox, and otherwise falls back to a written procedure (Section~\ref{sec:compiling-capsule}). \textbf{(3)~Execute}: In a later, warm execution, a plan executor
runs the programs and hands natural-language steps and failed programs to the model
(Section~\ref{sec:executing-capsule}). Each phase runs in a fresh sandbox, and no phase sees
the benchmark's tests or verifier (Appendix~\ref{app:sandboxes}).

These phases follow a tracing JIT compiler (Section~\ref{sec:background-jit}). The cold execution
is the recorded path, and the capsule is the compiled trace, kept for later requests of the same kind
as a trace cache keeps compiled traces~\citep{gal2009tracemonkey}. SkillCache's compiler is an LLM agent that can write failing programs, which is
why the host replays each candidate before admitting it. A program can still fail later when
the request, the model's decisions, or the environment differ. Like a failed guard, a failed program
leaves the compiled path through deoptimization, but only for that step, and the rest of the plan stays
compiled.

%% file: compiling_capsule.tex
\section{Compiling a Capsule}
\label{sec:compiling-capsule}

\subsection{The Capsule Representation}
\label{sec:capsule-format}

The compiler emits a parameterized capsule
\begin{equation}
  z=(\mu,B,P,\pi,V),
  \label{eq:capsule}
\end{equation}
where $\mu$ describes when the capsule applies, $B$ holds the bindings (the named values that programs
read or write), $P$ contains programs, $\pi$ interleaves program calls with natural-language steps, and $V$ is an
optional verification of the postconditions that the request and source skill state. $V$ runs after $\pi$ and, like $\pi$, mixes
program calls with natural-language steps.

Each binding describes how to obtain its live value, from the request or workspace, a natural-language
step, or a program's output. Each program declares the bindings it reads and those it sets for later
programs, so values pass from one program to the next without the model. Programs are language-agnostic. Each has source files, a command,
and a syntax or build check, reads its inputs from a JSON file, and prints one JSON result or
exits nonzero with a diagnostic.

Alongside $z$, the compiler returns examples $E$, the values the bindings had in the observed
execution. They serve only to qualify the capsule (Section~\ref{sec:capsule-qualification}). The
host removes them when it mounts the capsule, so a warm execution never receives $E$.

\subsection{Trace-Guided Compilation}
\label{sec:trace-compiler}

The compiler reads $(q,S,\tau_{\mathrm{obs}})$ in a fresh sandbox that holds none of the task's
files. The compiler prompt and the
program interface are fixed across tasks, and the prompt (Appendix~\ref{app:compiler-artifacts})
specifies four specialization rules. \textbf{\emph{Path pruning}}: The compiler removes failed
exploration, keeps the skill's requirements, and adds no behavior that neither the skill nor
the trace shows. \textbf{\emph{Binding lifting}}: Values that a program needs and that depend on the
request or workspace become bindings, each with an instruction for obtaining its live value, while
other values stay as the execution used them. \textbf{\emph{Region fusion}}: Deterministic operations
become programs, and adjacent ones share one program. Where the execution ran code, such as a
skill's script, the program reuses that code and changes only the bindings. A step that needs
semantic judgment or a new observation stays a natural-language step.
\textbf{\emph{Grounded postconditions}}: $V$ checks only the observable postconditions that the
request and source skill state, together with any inspection by which the execution checked and
corrected its own result. The prompt forbids comparing a result with the answer the observed
execution produced, and the compiler never sees the evaluator.

\subsection{Capsule Qualification}
\label{sec:capsule-qualification}

To reduce program failures at reuse, where recovery is costly (Section~\ref{sec:timing}), the host
qualifies each candidate before admitting it to the cache, through three gates that do not call the model.

\textbf{Gate 1: Structure and dataflow.}
The host parses the capsule and checks its dataflow. Steps call exactly the programs in $P$, and the
programs' inputs and outputs are exactly the bindings in $B$. Each output comes from only one program,
which runs before any program that reads it.

\textbf{Gate 2: Program checks.}
The host creates a fresh sandbox from the task image and runs the check the compiler wrote
for each program, such as \texttt{python3 -m py\_compile}. This gate catches programs that do not parse
or build in the task's environment.

\textbf{Gate 3: End-to-end replay.}
In a fresh task sandbox, the host replays the plan with the same plan executor used at reuse.
The examples $E$ supply every binding that no program outputs. Programs execute on task inputs and pass their outputs through the bindings. If a
program fails, replay records its diagnostics, substitutes its observed outputs from $E$, and
continues to collect failures from later programs in the same replay.

The compiler receives the collected failures and its previous answer, then returns a repaired
capsule. The host allows up to three attempts at the \emph{hybrid} form, a capsule with programs. If no
candidate qualifies, the host compiles the same trace into a written procedure, the \emph{textual}
form, with a prompt that follows the same rules. If that also fails, later executions run with the source skill alone.
Appendix~\ref{app:compiler-artifacts} gives the prompts for both forms.

%% file: executing_capsule.tex
\section{Executing a Capsule}
\label{sec:executing-capsule}

\subsection{The Plan Executor}
\label{sec:plan-executor}

\begin{figure}[!t]
  \centering
  \includegraphics[width=\columnwidth]{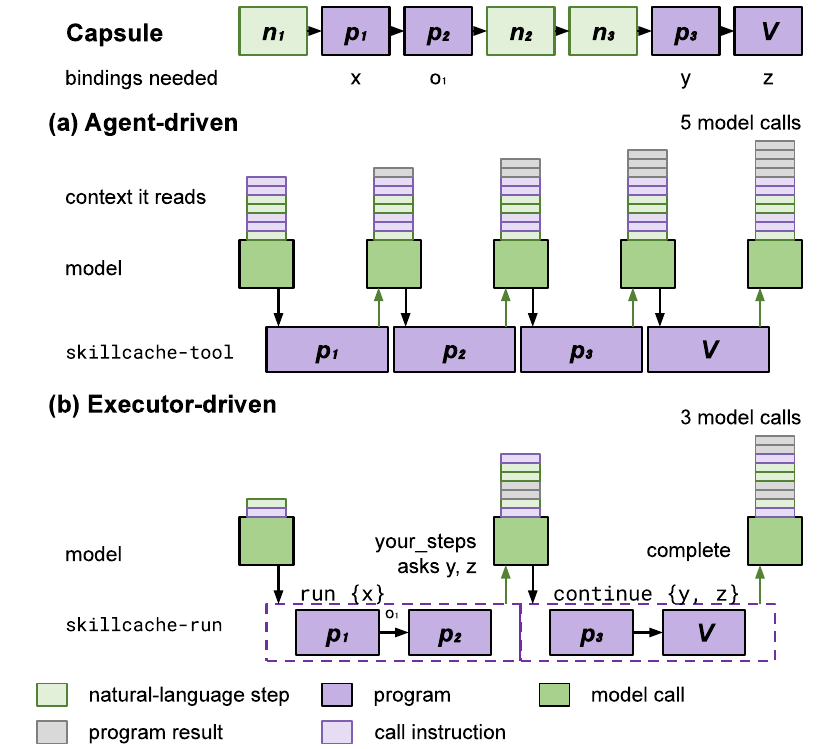}
  \caption{Who drives the plan. (a) Agent-driven: the model calls every program itself.
  (b) Executor-driven: the executor runs programs back to back and returns to the model only when
  the plan needs it. Bars illustrate the plan content each call reads beyond the shared context.}
  \label{fig:who-drives}

  \vspace{\floatsep}
  \includegraphics[width=\columnwidth]{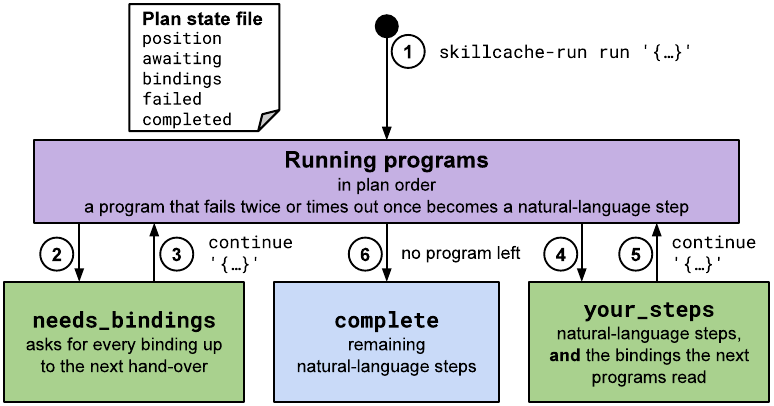}
  \caption{The plan executor as a state machine. Each \texttt{run} or \texttt{continue} is a short
  process that runs programs until the model is needed, reports, and exits. The plan state file
  carries the plan between calls. Circled numbers follow the Example
  paragraph.}
  \label{fig:executor-states}
\end{figure}

If the model called each program itself, it would still read the whole plan from the start and orchestrate every call. A capsule, however, explicitly interleaves program
steps with natural-language steps and declares each program's input and output bindings. This structure lets
consecutive programs run back to back without invoking the model. We thus introduce an \emph{executor-driven} design, in which a plan executor, \texttt{skillcache-run},
advances through the program steps and returns control to the model
when the plan needs its input or judgment. The executor runs each program through the runtime,
\texttt{skillcache-tool}. Both \texttt{skillcache-run} and
\texttt{skillcache-tool} are command-line programs that the host places in the agent's sandbox, and
the model invokes them with its ordinary shell tool, so the agent harness runs unchanged. In the alternative
\emph{agent-driven} backend, the prompt holds the whole plan and the model calls
\texttt{skillcache-tool} for each program itself (Figure~\ref{fig:who-drives}).

\textbf{Protocol.}
The prompt shows the capsule's name and description, and the model decides whether it applies. On a
miss, the model serves the request with the skills and its normal tools, as in a cold execution. On a
hit, the model first performs any steps before the first program, such as locating an input, and
then starts the executor once with \texttt{skillcache-run run} and the bindings that the programs read before the
first hand-over. A textual capsule has no programs, so the model follows its steps directly. The
executor runs program steps in order and returns to the model with one of three
reports (Figure~\ref{fig:executor-states}), each listing the programs completed since the previous report with
their results. \texttt{needs\_bindings} asks at once for every binding that the programs up to the next
hand-over read but were not given.
\texttt{your\_steps} hands over the next natural-language steps, including any step whose program failed
(Section~\ref{sec:deoptimization}), and asks for the bindings that the next programs read.
\texttt{complete} reports that no program is left, with the remaining
natural-language steps for the model. The model answers the first two with \texttt{skillcache-run continue} and the bindings asked
for.

\textbf{Plan state.}
Each \texttt{run} or \texttt{continue} is a short process that prints its report and exits. A plan
state file records the next step to run (\texttt{position}), the step a \texttt{continue} resumes at
(\texttt{awaiting}), every binding known so far (\texttt{bindings}), the steps whose program
failed with their diagnostics (\texttt{failed}), and the programs completed since the last report with their
results (\texttt{completed}), as Figure~\ref{fig:executor-states} lists. The executor saves it before each program runs, so when the agent's
command limit ends a call in the middle of a plan, a later \texttt{continue} reruns the interrupted
program.

\textbf{Fewer and smaller calls.}
The executor passes each program's outputs to the programs that read them and asks
at once for the other bindings that the programs up to the next hand-over read. Thus, the programs
run back to back without returning to the model for an input. When several natural-language steps come in a row, the executor
hands them over in one report, and the model performs them all before a single \texttt{continue}
with the bindings asked for. Each call also reads less of the plan, since later steps reach the model only when the
executor hands them over.

\textbf{Example.}
Consider the capsule at the top of Figure~\ref{fig:who-drives} and follow it through 
Figure~\ref{fig:executor-states}. The
model first performs $n_1$. Suppose it then starts the plan with \texttt{run} but leaves out $x$, which the
prompt lists (\textcircled{\scriptsize 1}). Program $p_1$ reads $x$, so the executor reports
\texttt{needs\_bindings} and asks for it (\textcircled{\scriptsize 2}). The model gives $x$ with
\texttt{continue} (\textcircled{\scriptsize 3}). The executor runs $p_1$ and passes its output $o_1$
to $p_2$, and $p_2$ fails twice (Section~\ref{sec:deoptimization}). The executor reports \texttt{your\_steps} with $p_2$, marked
\texttt{program\_failed}, together with $n_2$ and $n_3$, and asks in the same report for $y$
and $z$ (\textcircled{\scriptsize 4}). The model performs the three steps and answers with one
\texttt{continue} that gives $y$ and $z$, all that $p_3$ and $V$ read, and $p_3$ and $V$ run
(\textcircled{\scriptsize 5}). No program is left, so the executor reports \texttt{complete}
(\textcircled{\scriptsize 6}), and the model replies to the request.

\subsection{Deoptimization}
\label{sec:deoptimization}

A compiled program can fail on a later request, for example when a service is down or an input
differs from the observed one. A program fails when it exits nonzero twice, runs out of time, or
returns without an output it declares. The executor then hands that step to the model as
\texttt{program\_failed}, with its instruction and diagnostics. For a step of $\pi$ (Equation~\ref{eq:capsule}), it also gives the
path of the program's source, which shows how the step was done. Only the failing step leaves the
compiled path, and the rest of the plan continues compiled. The model then performs that step itself,
with the source skills still available as in a cold execution. In the main runs, the executor hands back
a failed program in 11 of 481 warm executions of hybrid capsules (Appendix~\ref{app:task-types}).

%% file: experiments_results.tex
\section{Evaluation}
\label{sec:evaluation}

\begin{table*}[!t]
  \centering
  \caption{Cold execution against SkillCache on 87 SkillsBench tasks, by task type. Panels (a) to (c)
  give the mean per task and the percent change in the total from cold to warm, and panel (d) gives mean
  reward and its change in percentage points. Each cell is the mean over three runs, and each change gives its standard
  deviation across runs.}
  \label{tab:main}
  {\footnotesize
  \setlength{\tabcolsep}{0.8pt}
  \input{tables/category_table.tex}}
\end{table*}

We evaluate SkillCache on all 87 SkillsBench tasks~\citep{li2026skillsbench}. Section~\ref{sec:eval-main}
measures warm execution against cold execution, and Section~\ref{sec:eval-representations} compares SkillCache with
reuse baselines. Sections~\ref{sec:eval-transfer} and~\ref{sec:eval-crossquery} study
cross-model and cross-query transfer, respectively. Section~\ref{sec:eval-analysis} analyzes where the
savings come from and when compilation pays for itself, and Section~\ref{sec:eval-ablations}
conducts ablation studies.

\subsection{Setup}
\label{sec:eval-setup}

\textbf{Pipeline.}
We evaluate the cold-compile-warm pipeline with Claude Sonnet~4.6~\citep{anthropic2026sonnet46} and
Claude Opus~4.8~\citep{anthropic2026opus48}, which we call Sonnet and Opus below, at medium effort using
the Claude Code
harness~\citep{anthropic2025claudecode}, with a one-hour timeout and a fresh sandbox per execution
(Appendix~\ref{app:experiment-provenance}).
A cold execution serves the request with the skills alone, a compiler agent with the same model
compiles its trace, and a warm execution serves the same request again. The warm execution starts in a
fresh sandbox without the cold execution's outputs, and Section~\ref{sec:eval-crossquery} tests changed
requests, including ones the original answer fails. We run the whole pipeline
three times with each model, each from its own cold executions, and report the mean and standard
deviation across runs. Cross-model transfer and the replay ablation run three times, once with each run's
capsules, and cross-query transfer runs three times with the first run's capsules.

\textbf{Baselines.}
Besides the cold execution, we compare SkillCache with two reuse baselines from prior work. Each warm
context runs in the same harness with the same request and skills, and adds one form of the cold
execution. \emph{Raw trace} adds the trace itself, as systems that
reuse trajectories in context do~\citep{fang2026memp}. \emph{Written procedure} adds a textual procedure, the representation used in
workflow memory~\citep{wang2025awm} and plan caching~\citep{zhang2025apc}. For a fair comparison, the
same compiler writes it from the trace with the prompt of the capsule's textual form
(Appendix~\ref{app:compiler-artifacts}). Both baselines are agent-driven.
\emph{SkillCache} adds the capsule and runs it with the plan executor. To separate the representation
from the executor, Section~\ref{sec:eval-ablations} also runs the capsule agent-driven against both
baselines.

\textbf{Metrics.}
We report provider-reported tokens, which include cache reads and writes, list-price cost, model
calls, latency, and verifier reward. Reward is the task verifier's score in $[0,1]$, which is 1 when every
check passes, and a few tasks give partial credit. Latency is the span from a task's first model call to the end of its last,
so it includes the commands and programs between calls. A missing outcome scores zero reward, and tokens, cost, and calls cover the tasks on which every
context (cold included) recorded tokens and reward. Warm executions exclude the compile cost because a one-trace capsule is compiled after its
request is served. We report this cost separately and amortize it over reuses in
Section~\ref{sec:eval-analysis}. Because a few long tasks can dominate totals, we also count the tasks
on which one context is better or worse, taking each task's three-run mean.

\subsection{Main Results}
\label{sec:eval-main}

In Table~\ref{tab:main}, warm execution uses $80.9\pm3.6$\% fewer tokens in total than cold execution with
Sonnet~4.6 and $74.3\pm6.3$\% fewer with Opus~4.8. Appendix~\ref{app:token-components} shows that warm execution reduces every token component, including output tokens by 85\% and 80\%. In each run, it uses fewer tokens on 84,
86, and 85 of 87 tasks with Sonnet and on 85, 86, and 86 with Opus. It also reduces cost by 82\% and 75\%
and latency by 72\% and 69\%, and mean reward stays within run-to-run noise with both models (Appendix~\ref{app:variance}). Token savings hold
in every task type and vary most across runs in cybersecurity and finance, where a few long tasks dominate
(Appendix~\ref{app:task-types}).

\subsection{Comparison with Reuse Baselines}
\label{sec:eval-representations}

\begin{table}[t]
  \centering
  \caption{SkillCache (Ours) against the other warm contexts on 87 tasks. Upper rows give three-run
  means of total tokens (millions), total cost (dollars), and mean reward, and lower rows count the
  tasks on which Ours is better/worse. A star marks $p<0.005$ in a two-sided sign test.}
  \label{tab:representations}
  \footnotesize
  \setlength{\tabcolsep}{1pt}
  \input{tables/representations_table.tex}
\end{table}

SkillCache saves more than both reuse baselines (Table~\ref{tab:representations}), so its savings
come from more than reusing the cold execution. The raw trace, the written procedure, and SkillCache use 54\%, 67\%, and 81\% fewer tokens than
the cold execution with Sonnet and 49\%, 52\%, and 74\% fewer with Opus, and SkillCache uses the fewest
in every run
(Appendix~\ref{app:variance}). Table~\ref{tab:versus} also provides model calls, latency, and exact
$p$ values. None of these three contexts changes mean reward by more than 0.01 from the cold execution,
which is within run-to-run noise. Section~\ref{sec:eval-ablations} further shows that SkillCache's savings come from both the capsule representation
and the executor-driven design.

\subsection{Cross-Model Transfer}
\label{sec:eval-transfer}

\begin{figure}[t]
  \centering
  \includegraphics[width=\columnwidth]{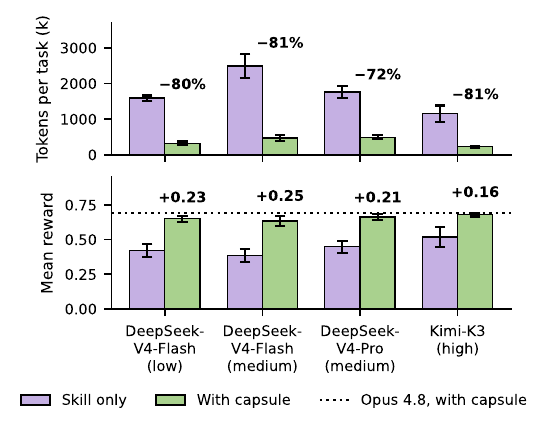}
  \caption{Cross-model transfer of Opus capsules on 87 tasks. Bars give three-run means with standard
  deviations, and the dotted line marks Opus's own reward.}
  \label{fig:transfer}
\end{figure}

\textbf{Setup.}
If capsules transfer across models, a strong model can compile once and an open-weight model can serve
recurring requests. To test this, we run DeepSeek-V4-Flash and
DeepSeek-V4-Pro~\citep{deepseek2026v4}, and Kimi-K3~\citep{kimi2026k3}, with the skills alone and
with the unchanged capsules of each Opus run. DeepSeek-V4-Flash runs at low and medium effort,
DeepSeek-V4-Pro at medium, and Kimi-K3 at high.

\textbf{Results.}
Figure~\ref{fig:transfer} shows that Opus capsules raise the mean reward of every model by 0.16 to
0.25, bringing it close to Opus's own reward of 0.690 with the same capsules, while using 72 to 81\%
fewer tokens than the skills alone. DeepSeek-V4-Flash shows similar gains at both effort levels, and the
token and reward changes are both significant for every model, with $p<0.005$ in a task-by-task sign
test. Appendix~\ref{app:versus} gives each model's numbers. We also run DeepSeek-V4-Flash at low effort with the Sonnet capsules and find a similar gain, with mean reward
rising from 0.437 to 0.553, against 0.546 for Sonnet, and 81\% fewer tokens.

\begin{table}[t]
  \centering
  \caption{Cross-query transfer to new requests. Each probe changes one thing in a source task and
  runs with the skills alone and with the source task's capsule. Cells are three-run means, and each
  change gives its standard deviation across runs.}
  \label{tab:crossquery}
  \footnotesize
  \setlength{\tabcolsep}{2pt}
  \input{tables/crossquery_table.tex}
\end{table}

\subsection{Cross-Query Transfer}
\label{sec:eval-crossquery}

\textbf{Setup.}
Because SkillsBench gives each task a single request, we additionally build three kinds of probes, each changing
one thing in a source task, to test whether a capsule still works when a recurring request changes. An
\emph{instance} probe changes the input content or a parameter the
request states, a \emph{rename} probe changes the paths of files the request names, and a
\emph{reword} probe rewrites one to three sentences of the request. A probe is kept only if, in its task's container,
its reference solution passes its verifier and, for an instance probe, the source task's reference
answer fails. We build 23 instance, 19 rename, and 24 reword probes for 24 of the 44
tasks whose cold execution succeeded in the first Sonnet run, all of which also succeeded with Opus
(Appendix~\ref{app:crossquery}).
We start from successful cold executions to test whether the resulting capsules can carry a successful
execution path over to a changed request.

\textbf{Results.}
Capsules stay effective when the request changes and keep their token savings
(Table~\ref{tab:crossquery}). Averaged over three
runs, tokens decrease by 70 to 87\% for every probe type with both models, and mean reward changes by
+0.01 to +0.15. On instance probes, where the original answer would fail, tokens decrease by 82\% with Sonnet and 72\%
with Opus while reward stays comparable. With DeepSeek-V4-Flash running the same Opus capsules, tokens
decrease by 71\% and mean reward rises from 0.54 to 0.92 across all probes (Appendix~\ref{app:crossquery}). One failure case comes from three Sonnet capsules that hard-code request-dependent values,
causing wrong answers on the corresponding instance probes. This highlights the need for generated programs
to use request-dependent values through bindings.

\subsection{Analysis}
\label{sec:eval-analysis}

\begin{table}[t]
  \centering
  \caption{Model calls and token savings by capsule type, pooled over the three runs, which hold 178, 65,
  and 18 task runs of the three kinds with Sonnet and 134, 113, and 14 with Opus.}
  \label{tab:kind}
  \footnotesize
  \setlength{\tabcolsep}{1.5pt}
  \input{tables/kind_table.tex}
\end{table}

\textbf{Model calls and savings by capsule type.}
Table~\ref{tab:kind} shows that savings follow model calls, since every call reads the
context again. In turn, the more of the plan programs cover, the fewer calls the model makes. When
every step is a program, the model makes about three calls per task. Natural-language steps and textual capsules leave more
of the work to the model, so they save less.

\textbf{Amortization.}
Compiling all 87 tasks costs \$67 with Sonnet and \$116 with Opus,
counting every attempt, repair, and textual fallback. One reuse saves \$64 and \$79, so
compilation is repaid after 1.1 reuses with Sonnet and 1.5 with Opus in aggregate, and after a median
of 1.3 and 2.2 reuses per task. In tokens, compilation and the first warm execution together
use 45\% fewer than the cold execution with Sonnet and 16\% fewer with Opus, so in aggregate one reuse
already covers the compilation cost. On average across runs, one reuse covers it on 67 of 87 tasks with Sonnet and 39
with Opus.

\subsection{Ablation Study}
\label{sec:eval-ablations}

\begin{table}[t]
  \centering
  \caption{Executor-driven against agent-driven execution of the same capsules on 87 tasks. Tokens are in millions and cost in dollars. A star marks $p<0.005$ in a two-sided sign test.}
  \label{tab:executor}
  \footnotesize
  \setlength{\tabcolsep}{1pt}
  \input{tables/executor_table.tex}
\end{table}

\textbf{Plan executor.}
Table~\ref{tab:executor} compares the executor-driven design with the agent-driven one on the same
capsules. The
executor-driven design uses 26\% and 21\% fewer tokens than the agent-driven one with Sonnet and
Opus, and fewer in each run (Appendix~\ref{app:variance}). It also needs fewer model calls on 77 and
70 of 87 tasks (Appendix~\ref{app:versus}), with no significant reward difference. Notably, even when all three are agent-driven, the capsule alone uses 44\% and 22\% fewer tokens than
the raw trace and the written procedure baselines with Sonnet and 38\% and 32\% fewer with Opus. We attribute the savings to the model seeing each program through its interface, which keeps
the code out of its context until needed, and to program steps that usually take a single call. Thus, both the capsule representation and the executor-driven design contribute to the
savings.

\begin{table}[t]
  \centering
  \caption{Replay against no replay, pooled over the three runs, on the task runs where replay changed the
  published capsule. No replay reuses the first candidate that passed the first two gates. Tokens are in millions and cost in
  dollars. A star marks $p<0.005$ in a two-sided sign test.}
  \label{tab:validation}
  \footnotesize
  \setlength{\tabcolsep}{1pt}
  \input{tables/validation_table.tex}
\end{table}

\textbf{Replay.}
Table~\ref{tab:validation} shows the effect of replay during qualification. To isolate it, we rerun each task
where replay changed the published capsule in its run, this time with the first candidate that passed
the first two gates (65 task runs with Sonnet and 79 with Opus, pooled over the three runs). Against that
candidate, the published capsule uses $59\pm28$\% fewer tokens with Sonnet and $36\pm7$\% fewer with Opus,
averaged over the three runs, fewer in every run (Appendix~\ref{app:variance}), and fewer on 56 of 65 and 66 of 79 task runs. Mean reward is 0.406 against 0.368
with Sonnet and 0.625 against 0.597 with Opus. Neither paired sign test is significant.

\begin{figure}[t]
  \centering
  \includegraphics[width=\columnwidth]{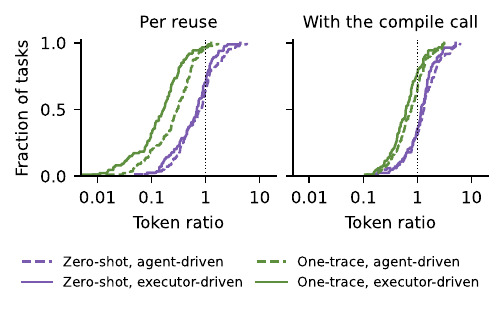}
  \caption{Zero-shot and first-run one-trace capsules with Sonnet~4.6, each executed agent-driven and
  executor-driven. Ratios are to the skills alone for zero-shot and to the cold execution for
  one-trace, and the right panel adds the compile call.}
  \label{fig:preexecution-cost-boundary}
\end{figure}

\textbf{Zero-shot with the executor.}
Section~\ref{sec:preexecution-study} runs zero-shot capsules agent-driven. To test whether the executor
alone closes their gap to one-trace capsules, we also run them with the plan executor and find that the gap
remains (Figure~\ref{fig:preexecution-cost-boundary}). Per reuse, executor-driven zero-shot capsules use fewer
tokens than the skills alone on 59 of 84 tasks, against 84 of 87 for first-run one-trace capsules. With the
compile call, they use fewer on 30 of 84 tasks, against 67 of 87 for one-trace capsules.

%% file: tables/category_table.tex
\begin{minipage}[t]{0.49\linewidth}
\centering
\textbf{(a) Tokens per task (thousands)}\par\vspace{1mm}
\begin{tabular*}{\linewidth}{@{\extracolsep{\fill}}lrr@{\extracolsep{0pt}\,}l@{\extracolsep{\fill}\hspace{1pt}}rr@{\extracolsep{0pt}\,}l@{}}
\toprule
 & \multicolumn{3}{c}{Sonnet 4.6} & \multicolumn{3}{c}{Opus 4.8} \\
\cmidrule(lr){2-4}\cmidrule(l){5-7}
Task type & Cold & \multicolumn{2}{c}{Warm} & Cold & \multicolumn{2}{c}{Warm} \\
\midrule
\textbf{Overall} & \textbf{1313} & \textbf{247} & \textbf{\textcolor{green!50!black}{{\scriptsize (\ensuremath{-}80.9{\tiny$\pm$3.6}\%)}}} & \textbf{1063} & \textbf{273} & \textbf{\textcolor{green!50!black}{{\scriptsize (\ensuremath{-}74.3{\tiny$\pm$6.3}\%)}}} \\
\midrule
Cybersecurity & 2020 & 579 & \textcolor{green!50!black}{{\scriptsize (\ensuremath{-}68.9{\tiny$\pm$35.4}\%)}} & 1264 & 526 & \textcolor{green!50!black}{{\scriptsize (\ensuremath{-}58.1{\tiny$\pm$17.0}\%)}} \\
Finance/econ. & 2382 & 651 & \textcolor{green!50!black}{{\scriptsize (\ensuremath{-}64.4{\tiny$\pm$28.4}\%)}} & 1402 & 326 & \textcolor{green!50!black}{{\scriptsize (\ensuremath{-}76.2{\tiny$\pm$14.9}\%)}} \\
Industrial/phys. & 1141 & 103 & \textcolor{green!50!black}{{\scriptsize (\ensuremath{-}91.0{\tiny$\pm$0.9}\%)}} & 977 & 203 & \textcolor{green!50!black}{{\scriptsize (\ensuremath{-}79.0{\tiny$\pm$4.4}\%)}} \\
Math/formal & 1165 & 135 & \textcolor{green!50!black}{{\scriptsize (\ensuremath{-}88.2{\tiny$\pm$2.4}\%)}} & 1589 & 186 & \textcolor{green!50!black}{{\scriptsize (\ensuremath{-}88.3{\tiny$\pm$1.5}\%)}} \\
Media & 559 & 101 & \textcolor{green!50!black}{{\scriptsize (\ensuremath{-}82.0{\tiny$\pm$9.5}\%)}} & 506 & 136 & \textcolor{green!50!black}{{\scriptsize (\ensuremath{-}72.6{\tiny$\pm$6.3}\%)}} \\
Natural sci. & 595 & 104 & \textcolor{green!50!black}{{\scriptsize (\ensuremath{-}82.6{\tiny$\pm$4.1}\%)}} & 466 & 140 & \textcolor{green!50!black}{{\scriptsize (\ensuremath{-}70.1{\tiny$\pm$3.2}\%)}} \\
Office & 833 & 182 & \textcolor{green!50!black}{{\scriptsize (\ensuremath{-}77.2{\tiny$\pm$9.9}\%)}} & 846 & 313 & \textcolor{green!50!black}{{\scriptsize (\ensuremath{-}63.1{\tiny$\pm$8.6}\%)}} \\
Software eng. & 1909 & 283 & \textcolor{green!50!black}{{\scriptsize (\ensuremath{-}84.7{\tiny$\pm$3.7}\%)}} & 1482 & 358 & \textcolor{green!50!black}{{\scriptsize (\ensuremath{-}75.8{\tiny$\pm$2.5}\%)}} \\
\bottomrule
\end{tabular*}
\end{minipage}
\hfill
\begin{minipage}[t]{0.49\linewidth}
\centering
\textbf{(b) Cost per task (\$)}\par\vspace{1mm}
\begin{tabular*}{\linewidth}{@{\extracolsep{\fill}}lrr@{\extracolsep{0pt}\,}l@{\extracolsep{\fill}\hspace{1pt}}rr@{\extracolsep{0pt}\,}l@{}}
\toprule
 & \multicolumn{3}{c}{Sonnet 4.6} & \multicolumn{3}{c}{Opus 4.8} \\
\cmidrule(lr){2-4}\cmidrule(l){5-7}
Task type & Cold & \multicolumn{2}{c}{Warm} & Cold & \multicolumn{2}{c}{Warm} \\
\midrule
\textbf{Overall} & \textbf{0.90} & \textbf{0.17} & \textbf{\textcolor{green!50!black}{{\scriptsize (\ensuremath{-}81.6{\tiny$\pm$1.0}\%)}}} & \textbf{1.21} & \textbf{0.30} & \textbf{\textcolor{green!50!black}{{\scriptsize (\ensuremath{-}75.1{\tiny$\pm$5.5}\%)}}} \\
\midrule
Cybersecurity & 1.24 & 0.40 & \textcolor{green!50!black}{{\scriptsize (\ensuremath{-}67.8{\tiny$\pm$34.6}\%)}} & 1.23 & 0.57 & \textcolor{green!50!black}{{\scriptsize (\ensuremath{-}53.7{\tiny$\pm$22.0}\%)}} \\
Finance/econ. & 1.62 & 0.41 & \textcolor{green!50!black}{{\scriptsize (\ensuremath{-}71.1{\tiny$\pm$22.0}\%)}} & 1.94 & 0.34 & \textcolor{green!50!black}{{\scriptsize (\ensuremath{-}82.3{\tiny$\pm$8.9}\%)}} \\
Industrial/phys. & 1.02 & 0.07 & \textcolor{green!50!black}{{\scriptsize (\ensuremath{-}93.1{\tiny$\pm$1.0}\%)}} & 1.35 & 0.32 & \textcolor{green!50!black}{{\scriptsize (\ensuremath{-}75.9{\tiny$\pm$3.1}\%)}} \\
Math/formal & 1.08 & 0.12 & \textcolor{green!50!black}{{\scriptsize (\ensuremath{-}88.2{\tiny$\pm$2.3}\%)}} & 1.74 & 0.23 & \textcolor{green!50!black}{{\scriptsize (\ensuremath{-}87.0{\tiny$\pm$2.9}\%)}} \\
Media & 0.26 & 0.05 & \textcolor{green!50!black}{{\scriptsize (\ensuremath{-}79.6{\tiny$\pm$10.2}\%)}} & 0.45 & 0.13 & \textcolor{green!50!black}{{\scriptsize (\ensuremath{-}70.4{\tiny$\pm$6.2}\%)}} \\
Natural sci. & 0.42 & 0.06 & \textcolor{green!50!black}{{\scriptsize (\ensuremath{-}85.2{\tiny$\pm$4.8}\%)}} & 0.52 & 0.13 & \textcolor{green!50!black}{{\scriptsize (\ensuremath{-}74.2{\tiny$\pm$1.6}\%)}} \\
Office & 0.63 & 0.13 & \textcolor{green!50!black}{{\scriptsize (\ensuremath{-}79.3{\tiny$\pm$6.0}\%)}} & 0.85 & 0.33 & \textcolor{green!50!black}{{\scriptsize (\ensuremath{-}60.6{\tiny$\pm$5.4}\%)}} \\
Software eng. & 1.02 & 0.19 & \textcolor{green!50!black}{{\scriptsize (\ensuremath{-}81.3{\tiny$\pm$0.9}\%)}} & 1.54 & 0.35 & \textcolor{green!50!black}{{\scriptsize (\ensuremath{-}77.6{\tiny$\pm$2.1}\%)}} \\
\bottomrule
\end{tabular*}
\end{minipage}
\par\vspace{2mm}
\begin{minipage}[t]{0.49\linewidth}
\centering
\textbf{(c) Latency per task (s)}\par\vspace{1mm}
\begin{tabular*}{\linewidth}{@{\extracolsep{\fill}}lrr@{\extracolsep{0pt}\,}l@{\extracolsep{\fill}\hspace{1pt}}rr@{\extracolsep{0pt}\,}l@{}}
\toprule
 & \multicolumn{3}{c}{Sonnet 4.6} & \multicolumn{3}{c}{Opus 4.8} \\
\cmidrule(lr){2-4}\cmidrule(l){5-7}
Task type & Cold & \multicolumn{2}{c}{Warm} & Cold & \multicolumn{2}{c}{Warm} \\
\midrule
\textbf{Overall} & \textbf{372} & \textbf{103} & \textbf{\textcolor{green!50!black}{{\scriptsize (\ensuremath{-}72.1{\tiny$\pm$2.4}\%)}}} & \textbf{373} & \textbf{113} & \textbf{\textcolor{green!50!black}{{\scriptsize (\ensuremath{-}69.2{\tiny$\pm$7.2}\%)}}} \\
\midrule
Cybersecurity & 387 & 165 & \textcolor{green!50!black}{{\scriptsize (\ensuremath{-}57.3{\tiny$\pm$13.1}\%)}} & 278 & 178 & \textcolor{green!50!black}{{\scriptsize (\ensuremath{-}35.3{\tiny$\pm$11.2}\%)}} \\
Finance/econ. & 468 & 125 & \textcolor{green!50!black}{{\scriptsize (\ensuremath{-}70.5{\tiny$\pm$20.9}\%)}} & 466 & 75 & \textcolor{green!50!black}{{\scriptsize (\ensuremath{-}80.3{\tiny$\pm$14.8}\%)}} \\
Industrial/phys. & 452 & 60 & \textcolor{green!50!black}{{\scriptsize (\ensuremath{-}86.3{\tiny$\pm$2.9}\%)}} & 439 & 123 & \textcolor{green!50!black}{{\scriptsize (\ensuremath{-}72.0{\tiny$\pm$1.5}\%)}} \\
Math/formal & 636 & 216 & \textcolor{green!50!black}{{\scriptsize (\ensuremath{-}64.1{\tiny$\pm$16.1}\%)}} & 843 & 198 & \textcolor{green!50!black}{{\scriptsize (\ensuremath{-}73.3{\tiny$\pm$19.5}\%)}} \\
Media & 84 & 36 & \textcolor{green!50!black}{{\scriptsize (\ensuremath{-}57.2{\tiny$\pm$13.5}\%)}} & 129 & 60 & \textcolor{green!50!black}{{\scriptsize (\ensuremath{-}50.4{\tiny$\pm$18.8}\%)}} \\
Natural sci. & 343 & 132 & \textcolor{green!50!black}{{\scriptsize (\ensuremath{-}57.8{\tiny$\pm$14.7}\%)}} & 271 & 128 & \textcolor{green!50!black}{{\scriptsize (\ensuremath{-}53.0{\tiny$\pm$5.5}\%)}} \\
Office & 258 & 55 & \textcolor{green!50!black}{{\scriptsize (\ensuremath{-}74.7{\tiny$\pm$12.4}\%)}} & 186 & 70 & \textcolor{green!50!black}{{\scriptsize (\ensuremath{-}62.3{\tiny$\pm$4.9}\%)}} \\
Software eng. & 320 & 86 & \textcolor{green!50!black}{{\scriptsize (\ensuremath{-}73.8{\tiny$\pm$8.0}\%)}} & 416 & 97 & \textcolor{green!50!black}{{\scriptsize (\ensuremath{-}76.6{\tiny$\pm$2.7}\%)}} \\
\bottomrule
\end{tabular*}
\end{minipage}
\hfill
\begin{minipage}[t]{0.49\linewidth}
\centering
\textbf{(d) Reward}\par\vspace{1mm}
\begin{tabular*}{\linewidth}{@{\extracolsep{\fill}}lrr@{\extracolsep{0pt}\,}l@{\extracolsep{\fill}\hspace{1pt}}rr@{\extracolsep{0pt}\,}l@{}}
\toprule
 & \multicolumn{3}{c}{Sonnet 4.6} & \multicolumn{3}{c}{Opus 4.8} \\
\cmidrule(lr){2-4}\cmidrule(l){5-7}
Task type & Cold & \multicolumn{2}{c}{Warm} & Cold & \multicolumn{2}{c}{Warm} \\
\midrule
\textbf{Overall} & \textbf{0.538} & \textbf{0.546} & \textbf{\textcolor{green!50!black}{{\scriptsize (\ensuremath{+}0.8{\tiny$\pm$2.7} pp)}}} & \textbf{0.692} & \textbf{0.690} & \textbf{\textcolor{red!65!black}{{\scriptsize (\ensuremath{-}0.2{\tiny$\pm$1.1} pp)}}} \\
\midrule
Cybersecurity & 0.531 & 0.603 & \textcolor{green!50!black}{{\scriptsize (\ensuremath{+}7.1{\tiny$\pm$18.6} pp)}} & 0.634 & 0.610 & \textcolor{red!65!black}{{\scriptsize (\ensuremath{-}2.4{\tiny$\pm$2.4} pp)}} \\
Finance/econ. & 0.222 & 0.185 & \textcolor{red!65!black}{{\scriptsize (\ensuremath{-}3.7{\tiny$\pm$6.4} pp)}} & 0.444 & 0.444 & {\scriptsize (\ensuremath{+}0.0{\tiny$\pm$0.0} pp)} \\
Industrial/phys. & 0.492 & 0.492 & {\scriptsize (\ensuremath{+}0.0{\tiny$\pm$0.0} pp)} & 0.669 & 0.669 & {\scriptsize (\ensuremath{+}0.0{\tiny$\pm$7.1} pp)} \\
Math/formal & 0.481 & 0.510 & \textcolor{green!50!black}{{\scriptsize (\ensuremath{+}2.9{\tiny$\pm$5.1} pp)}} & 0.650 & 0.650 & {\scriptsize (\ensuremath{+}0.0{\tiny$\pm$0.0} pp)} \\
Media & 0.667 & 0.667 & {\scriptsize (\ensuremath{+}0.0{\tiny$\pm$0.0} pp)} & 0.733 & 0.733 & {\scriptsize (\ensuremath{+}0.0{\tiny$\pm$0.0} pp)} \\
Natural sci. & 0.808 & 0.808 & {\scriptsize (\ensuremath{+}0.0{\tiny$\pm$0.0} pp)} & 0.905 & 0.905 & {\scriptsize (\ensuremath{+}0.0{\tiny$\pm$0.0} pp)} \\
Office & 0.571 & 0.619 & \textcolor{green!50!black}{{\scriptsize (\ensuremath{+}4.8{\tiny$\pm$4.1} pp)}} & 0.738 & 0.738 & {\scriptsize (\ensuremath{+}0.0{\tiny$\pm$0.0} pp)} \\
Software eng. & 0.480 & 0.459 & \textcolor{red!65!black}{{\scriptsize (\ensuremath{-}2.1{\tiny$\pm$3.6} pp)}} & 0.657 & 0.657 & {\scriptsize (\ensuremath{+}0.0{\tiny$\pm$0.0} pp)} \\
\bottomrule
\end{tabular*}
\end{minipage}

%% file: tables/representations_table.tex
\begin{tabular*}{\linewidth}{@{\extracolsep{\fill}}l*{6}{c}@{}}
\toprule
 & \multicolumn{3}{c}{Sonnet~4.6} & \multicolumn{3}{c}{Opus~4.8} \\
\cmidrule(lr){2-4}\cmidrule(l){5-7}
Context & Tokens & Cost & Reward & Tokens & Cost & Reward \\
\midrule
\multicolumn{7}{@{}l}{\emph{Totals}} \\
Raw trace & 52.6 & 38.48 & 0.541 & 47.4 & 54.75 & \textbf{0.700} \\
Procedure & 37.6 & 27.82 & 0.541 & 44.2 & 47.41 & 0.685 \\
Ours & \textbf{21.5} & \textbf{14.41} & \textbf{0.546} & \textbf{23.7} & \textbf{26.16} & 0.690 \\
\midrule
\multicolumn{7}{@{}l}{\emph{Ours better/worse, task by task}} \\
vs raw trace & 77/10$^{*}$ & 78/9$^{*}$ & 5/3 & 75/12$^{*}$ & 78/9$^{*}$ & 0/4 \\
vs procedure & 82/5$^{*}$ & 82/5$^{*}$ & 5/4 & 82/5$^{*}$ & 79/8$^{*}$ & 3/4 \\
\bottomrule
\end{tabular*}

%% file: tables/crossquery_table.tex
\begin{tabular*}{\linewidth}{@{\extracolsep{\fill}}lrrrr@{}}
\toprule
 & \multicolumn{2}{c}{Tokens per probe} & \multicolumn{2}{c}{Mean reward} \\
\cmidrule(lr){2-3}\cmidrule(l){4-5}
Probe & Skill only & With capsule & Skill only & With capsule \\
\midrule
\multicolumn{5}{@{}l}{\emph{Sonnet~4.6}} \\
\textbf{Overall} & \textbf{658k} & \textbf{117k \textcolor{green!50!black}{{\scriptsize ($-$82{\tiny$\pm$2}\%)}}} & \textbf{0.81} & \textbf{0.89 \textcolor{green!50!black}{{\scriptsize (+0.08{\tiny$\pm$0.04})}}} \\
Instance & 660k & 117k \textcolor{green!50!black}{{\scriptsize ($-$82{\tiny$\pm$2}\%)}} & 0.81 & 0.83 \textcolor{green!50!black}{{\scriptsize (+0.01{\tiny$\pm$0.05})}} \\
Rename & 649k & 162k \textcolor{green!50!black}{{\scriptsize ($-$74{\tiny$\pm$8}\%)}} & 0.81 & 0.88 \textcolor{green!50!black}{{\scriptsize (+0.07{\tiny$\pm$0.03})}} \\
Reword & 663k & 83k \textcolor{green!50!black}{{\scriptsize ($-$87{\tiny$\pm$1}\%)}} & 0.81 & 0.96 \textcolor{green!50!black}{{\scriptsize (+0.15{\tiny$\pm$0.13})}} \\
\midrule
\multicolumn{5}{@{}l}{\emph{Opus~4.8}} \\
\textbf{Overall} & \textbf{688k} & \textbf{190k \textcolor{green!50!black}{{\scriptsize ($-$72{\tiny$\pm$1}\%)}}} & \textbf{0.91} & \textbf{0.97 \textcolor{green!50!black}{{\scriptsize (+0.06{\tiny$\pm$0.03})}}} \\
Instance & 782k & 219k \textcolor{green!50!black}{{\scriptsize ($-$72{\tiny$\pm$4}\%)}} & 0.88 & 0.93 \textcolor{green!50!black}{{\scriptsize (+0.05{\tiny$\pm$0.05})}} \\
Rename & 575k & 170k \textcolor{green!50!black}{{\scriptsize ($-$70{\tiny$\pm$3}\%)}} & 0.95 & 1.00 \textcolor{green!50!black}{{\scriptsize (+0.05{\tiny$\pm$0.05})}} \\
Reword & 688k & 178k \textcolor{green!50!black}{{\scriptsize ($-$74{\tiny$\pm$2}\%)}} & 0.92 & 1.00 \textcolor{green!50!black}{{\scriptsize (+0.08{\tiny$\pm$0.04})}} \\
\bottomrule
\end{tabular*}

%% file: tables/kind_table.tex
\begin{tabular}{@{}lrrrrrr@{}}
\toprule
 & \multicolumn{3}{c}{Sonnet~4.6} & \multicolumn{3}{c}{Opus~4.8} \\
\cmidrule(lr){2-4}\cmidrule(l){5-7}
 & \multicolumn{2}{c}{Model calls} & Tokens & \multicolumn{2}{c}{Model calls} & Tokens \\
\cmidrule(lr){2-3}\cmidrule(lr){5-6}
Capsule type & Cold & Warm & saved & Cold & Warm & saved \\
\midrule
All tasks & 23.6 & \textbf{5.6} & 81\% & 20.9 & \textbf{6.7} & 74\% \\
\midrule
\multicolumn{7}{@{}l}{\emph{Hybrid}} \\
Programs only & 18.8 & \textbf{3.0} & 89\% & 17.2 & \textbf{3.4} & 87\% \\
Natural-language steps & 31.6 & \textbf{8.6} & 78\% & 25.1 & \textbf{10.1} & 67\% \\
\midrule
Textual & 41.9 & \textbf{20.9} & 56\% & 22.6 & \textbf{12.0} & 48\% \\
\bottomrule
\end{tabular}

%% file: tables/executor_table.tex
\begin{tabular*}{\linewidth}{@{\extracolsep{\fill}}l*{6}{c}@{}}
\toprule
 & \multicolumn{3}{c}{Sonnet~4.6} & \multicolumn{3}{c}{Opus~4.8} \\
\cmidrule(lr){2-4}\cmidrule(l){5-7}
Context & Tokens & Cost & Reward & Tokens & Cost & Reward \\
\midrule
\multicolumn{7}{@{}l}{\emph{Totals}} \\
Agent-driven & 29.0 & 17.40 & 0.544 & 29.7 & 30.05 & 0.678 \\
Executor-driven & \textbf{21.5} & \textbf{14.41} & \textbf{0.546} & \textbf{23.7} & \textbf{26.16} & \textbf{0.690} \\
\midrule
\multicolumn{7}{@{}l}{\emph{Executor-driven better/worse, task by task}} \\
vs agent-driven & 78/9$^{*}$ & 78/9$^{*}$ & 4/4 & 70/14$^{*}$ & 67/17$^{*}$ & 3/0 \\
\bottomrule
\end{tabular*}

%% file: tables/validation_table.tex
\begin{tabular*}{\linewidth}{@{\extracolsep{\fill}}l*{6}{c}@{}}
\toprule
 & \multicolumn{3}{c}{Sonnet~4.6 ($n=65$)} & \multicolumn{3}{c}{Opus~4.8 ($n=79$)} \\
\cmidrule(lr){2-4}\cmidrule(l){5-7}
Context & Tokens & Cost & Reward & Tokens & Cost & Reward \\
\midrule
\multicolumn{7}{@{}l}{\emph{Totals}} \\
No replay & 80.4 & 48.23 & 0.368 & 48.7 & 52.20 & 0.597 \\
Replay & \textbf{28.2} & \textbf{19.73} & \textbf{0.406} & \textbf{31.8} & \textbf{35.10} & \textbf{0.625} \\
\midrule
\multicolumn{7}{@{}l}{\emph{Replay better/worse, per task run}} \\
vs no replay & 56/9$^{*}$ & 53/12$^{*}$ & 5/3 & 66/13$^{*}$ & 67/12$^{*}$ & 3/0 \\
\bottomrule
\end{tabular*}

%% file: limitations.tex
\section{Limitations and Future Work}
\label{sec:limitations}

In our study, all experiments use SkillsBench and the Claude Code harness. We see broader potential in
the idea of compiling a completed execution into programs that an executor drives on later
requests. It could also benefit recurring agent tasks without skills or in other harnesses. Extending SkillCache to these settings is one direction for future work.

SkillCache also incurs a one-time compilation cost and is most useful when later requests amortize it
(Section~\ref{sec:eval-analysis}). In addition, each capsule is compiled from one cold execution and
may carry useful patterns alongside brittle assumptions, errors, or biases. Replay in qualification also runs
the programs only with the original request's values and inputs, so a capsule that hard-codes one of
those values can still pass, as the instance probes in Section~\ref{sec:eval-crossquery} show. Future work could let a capsule learn from its warm executions and improve
with use, much as a tracing JIT compiles paths it discovers later~\citep{gal2009tracemonkey}.

Finally, as SkillsBench supplies each task's skills, our evaluation supplies each task's capsule and does not
test retrieval.
Production use would require reliable retrieval and selection as the cache grows, along with execution
safeguards and cache invalidation when a skill or its environment changes. Integrating these
lifecycle and security mechanisms in persistent deployments is another direction for future work.

%% file: related_work.tex
\section{Related Work}
\label{sec:related}

\textbf{Skill evolution and source-level compilation.}
Skill-evolution methods construct or refine skills using rewarded trajectories
\citep{mi2026skillpro,xia2026skillrl,yang2026skillmaster}, held-out
rollouts~\citep{yang2026skillopt,tang2026wikiskill}, or other execution feedback
\citep{ni2026trace2skill,zhang2026coevoskills,zhou2026spark,zhang2026skillevolver}. A separate line of work compresses skill content~\citep{gao2026skillreducer}
or translates skill packages into typed representations~\citep{ouyang2026skcc}, runtime
interfaces~\citep{xu2026skillsmith}, and executable harnesses~\citep{dantanarayana2026sigil}.
SkillCache complements both by preserving the supplied skill and separately storing a
specialization compiled from one completed skill-guided invocation.

\textbf{Execution-grounded specialization.}
Existing systems learn code routines, executable skills, or workflows from interactions,
demonstrations, and traces
\citep{wang2024voyager,wang2025asi,zheng2025skillweaver,chen2026skillcraft,wang2026webxskill,hattami2026skillnb,guo2026skilldisco,shao2026nsi}.
SkVM~\citep{chen2026skvm} compiles skills for different
model-harness targets and promotes code patterns after repeated signature matches across
invocations. TraceCompiler~\citep{elyadouni2026tracecompiler} compiles clusters of agent traces
into mostly deterministic workflows. SkillTracer~\citep{li2026skilltracer} localizes failures within skill plan graphs and repairs
the affected structure.
When external evaluation is available, PreAct~\citep{li2026preact} compiles one successful GUI trace
into a gated state machine, and SkillDroid~\citep{chen2026skilldroid} compiles such a trace into an
action template replayed without per-step model calls. Other work likewise validates artifacts with verifier
feedback~\citep{rao2026skillopt} or deterministic checks~\citep{lu2026contractskill}. SkillCache instead specializes one
completed skill-guided invocation without benchmark reward or verifier access and retains the source skill for
recovery.

\textbf{Planning, program compilation, and harness optimization.}
ReWOO~\citep{xu2023rewoo} and Plan-and-Act~\citep{erdogan2025planandact} separate planning from
execution, LLMCompiler~\citep{kim2024llmcompiler} parallelizes function calls while respecting task
dependencies, and Agent JIT Compilation~\citep{winston2026agentjit} turns web-task descriptions into
cost-ranked code plans before execution. DSPy~\citep{khattab2024dspy} compiles LM programs, Meta-Harness~\citep{lee2026metaharness} searches
over harness code, and SoL-Pi~\citep{liu2026solpirecursivelyscalingautoresearch} searches for
token-reducing harness mechanisms, among them fusing a file edit and its follow-up command into one
tool call. SkillCache instead specializes a
supplied skill from one completed invocation into a capsule whose programs take whole runs of
deterministic tool calls away from the model. Unlike LLM-as-Code~\citep{qi2026llmascode}, where a program
also owns control flow and calls the model only for reasoning, our executor drives the capsule inside an unchanged agent harness.

\textbf{Procedural reuse and inference caching.}
Prior systems reuse model-readable insights, workflows, or procedural memories
\citep{zhao2024expel,wang2025awm,ouyang2026reasoningbank,fang2026memp}, cache plan templates and
transitions~\citep{zhang2025apc,kim2026agenticcache}, build reusable tools~\citep{cai2024latm}, or
synthesize executable response generators~\citep{chakraborty2025gencache}. At the inference layer, Prompt Cache~\citep{gim2024promptcache}
reuses attention states across prompts, while Don't Break the
Cache~\citep{lumer2026dontbreak} evaluates exact-prefix prompt
caching for agentic workloads. SkillCache instead caches programs and natural-language steps derived from a
skill-guided trace.

%% file: conclusion.tex
\section{Conclusion}
\label{sec:conclusion}

We study how repeated skill orchestration can be compiled into reusable execution. Compiling from the
skill and request alone does not reliably reduce tokens, so SkillCache
grounds each capsule in one completed execution, replays it before reuse, and drives it with a plan
executor, keeping the source skill for recovery. Warm executions use 81\% and 74\% fewer tokens with Sonnet and Opus at comparable
reward, and capsules stay effective under controlled request changes and transfer across models.

%% file: ai_usage_disclosure.tex
\section*{AI Usage Disclosure}
We used LLM-based coding assistants, Anthropic Claude Code and OpenAI Codex, in this work. The authors
conceived the research question and designed the method and the experiments, and discussed
with the assistants how to refine them. The authors searched for and read the
related literature, and used the assistants to make the search more complete and to summarize papers. The assistants supported method implementation (writing and refactoring code under the authors' direction),
experiment execution (automating and monitoring runs), result analysis (writing scripts that aggregate
results, build tables, and plot figures, and examining the results), and the generation of the cross-query
probes from SkillsBench tasks. For the manuscript, the authors discussed the wording with the assistants,
and the assistants helped with editing, polishing, LaTeX formatting, and drawing the method diagrams.
The language models evaluated in Section~\ref{sec:evaluation} serve as the SkillCache compiler
and task executors and are the object of study. All reported numbers come from benchmark executions
scored by the tasks' verifiers. The authors reviewed and verified AI-assisted work and take
full responsibility for this paper.

%% file: appendix.tex
\section{Experimental Details and Additional Results}
\label{app:additional-results}

\subsection{Experimental Details}
\label{app:experiment-provenance}
Every execution uses a fresh sandbox and a one-hour timeout, with at most 15 tasks in parallel, or 12
in zero-shot specialization and cross-model transfer. Matched conditions use the same provider defaults for
temperature and \texttt{top\_p}. Within a task, executions run one after another. The cold execution, and
the direct execution of the zero-shot study, run first, and the order of the others rotates across tasks
so that none always runs first. An attempt runs again when the harness fails before the agent starts or
when the agent produces no output for ten minutes, and executions that could not run as planned run
again in a later invocation. In both cases the last finished attempt counts, regardless of reward. The
first run of each model reuses cold executions recorded earlier under the same execution protocol,
which serve as both its cold baseline and the input its compiler reads. The raw trace
baseline receives the observable trajectory the compiler reads, as JSON, with each image's encoded bytes
replaced by a note, so that the baseline is not charged for base64 text that the cold model saw as an
image.

In cross-model transfer and the cross-query probes, token comparisons exclude task-run pairs with
missing or incomplete usage records, including seven cross-model pairs with partially recorded provider
usage. Rewards cover every task and probe. Because parallel execution introduces
contention, we treat latency as system-descriptive. Latency additionally excludes tasks if any compared
execution ends in an error or lacks timing records. Cost uses the published list prices per million
tokens, \$3 for input and \$15 for output with Sonnet~4.6 and \$5 and \$25 with Opus~4.8, with cache
reads at 0.1 times and five-minute cache writes at 1.25 times the input price. We access all models
through hosted APIs.

For environment compatibility, we patch three task images in every run. Two seismic images pin
\texttt{setuptools<81} to preserve the \texttt{pkg\_resources} interface required by their dependencies,
and the \texttt{civ6-adjacency-optimizer} image makes its output directory writable by the unprivileged agent
user. The same images are used across matched conditions, with no change to requests, skills,
fixtures, evaluators, or scoring. For the later cross-model runs, the \texttt{multilingual-video-dubbing}
image also pins four packages to their previous releases, because new releases broke its build.

\subsection{Zero-Shot Specialization}
\label{app:zero-shot}

This study runs once, with medium effort for direct execution, compilation, and capsule execution.
The compiler is the main study's compiler with Sonnet~4.6 and an empty trace. Its user message adds
that no execution has happened yet, that it should use only the request and the source skills without
inspecting the task workspace, and that it should package as programs only the deterministic operations
those specifications justify, leaving decisions that need a runtime observation as natural-language
steps. Its candidates go through the same syntax and build checks in a sandbox without task data, but
not replay, since no execution supplies observed values. After three failed attempts it falls back to the
textual form, which happened on 2 of the 87 tasks. Each capsule then runs agent-driven
(Section~\ref{sec:timing}) and executor-driven (Section~\ref{sec:eval-ablations}). We read each agent-driven trajectory to record whether a compiled program failed observably,
through an exception, a missing dependency, a timeout, or a bug in the program, and whether a
program returned a wrong answer without an error and that answer reached the final output. A program does not count as failed when its non-zero
exit came from an input the model supplied, from a check that correctly reported the model's own code,
or from stray lines on standard output beside correct results.

Table~\ref{tab:zero-shot-ratios} gives the per-task token ratios, total-token changes, and sign test
$p$ values of Section~\ref{sec:preexecution-study}. Each ratio is a geometric mean over tasks with a
95\% interval from 10,000 task-bootstrap resamples. One task holds 18\% of the direct tokens and alone
turns an 11\% increase in total tokens, compile call included, into a 5\% decrease. Counting execution
alone, zero-shot still uses more tokens than direct execution on 30 of the 84 tasks, where persisting
the capsule never repays compilation.

\begin{table*}[t]
  \centering
  \caption{Zero-shot specialization against direct execution on 84 tasks (Sonnet~4.6). Each cell gives
  the per-task token ratio, the change in total tokens, and the two-sided sign test $p$ value of the
  task counts. The first three rows are agent-driven (Section~\ref{sec:timing}), and the last is
  executor-driven (Section~\ref{sec:eval-ablations}). The No failure and Failure rows split the tasks by
  whether a compiled program failed observably. For agent-driven execution, mean reward decreases from 0.517 to 0.459,
  and reward rises on 10 tasks and decreases on 12 ($p=0.83$).}
  \label{tab:zero-shot-ratios}
  \footnotesize
  \input{tables/zero_shot_ratio_table.tex}
\end{table*}

\begin{table*}[t]
  \caption{What each phase receives. The compiler runs without the task's input files and sees them
  only through the trace and the failures that replay reports. In replay, the host gives the examples $E$ as bindings
  in place of the model. $^{*}$An agent-driven warm execution has no plan executor, and a textual
  capsule is given in the prompt with nothing mounted.}
  \label{tab:sandboxes}
  \centering
  \footnotesize
  \begin{tabularx}{\linewidth}{@{}Xccccc@{}}
    \toprule
    Receives & Cold & Compile & Check & Replay & Warm \\
    \midrule
    The model, with the agent's tools & $\checkmark$ & $\checkmark$ & & & $\checkmark$ \\
    The request & $\checkmark$ & $\checkmark$ & $\checkmark$ & $\checkmark$ & $\checkmark$ \\
    The source skills & $\checkmark$ & $\checkmark$ & & $\checkmark$ & $\checkmark$ \\
    The task's input files & $\checkmark$ & partly & $\checkmark$ & $\checkmark$ & $\checkmark$ \\
    The trace and the compile instruction & & $\checkmark$ & & & \\
    The runtime and the plan executor & & & $\checkmark$ & $\checkmark$ & $\checkmark^{*}$ \\
    The capsule without $E$, and its program sources & & & $\checkmark$ & $\checkmark$ & $\checkmark$ \\
    The examples $E$ & & & & $\checkmark$ & \\
    \bottomrule
  \end{tabularx}
\end{table*}

\subsection{Sandboxes per Phase}
\label{app:sandboxes}

Every phase runs in its own fresh sandbox (Table~\ref{tab:sandboxes}). No phase sees the
benchmark's tests, reference solutions (its oracle), verifier, or rewards. Every phase runs as the unprivileged agent
user, and those files are restricted to root. A verifier is uploaded only after a cold or warm
execution has finished, for scoring, so it never guides compilation or triggers recovery. The trace
the compiler reads excludes provider reasoning and redacts credentials. Programs run in the
agent's starting directory and read the source skills' files from a fixed read-only path.
Environment variables named like keys, tokens, or passwords are removed from their environment.

\subsection{Token Components}
\label{app:token-components}

Table~\ref{tab:components} decomposes the tokens of Table~\ref{tab:main} by provider component. Cache reads
make up about 95\% of the cold tokens with both models. SkillCache reduces every component, including output
tokens by 85\% and 80\%, so its savings do not come only from cache reads.

\begin{table}[t]
  \centering
  \caption{Cold and warm tokens by provider component, as in Table~\ref{tab:main}, in millions
  and averaged over the three runs. Reduction is the mean of each run's reduction, with its standard
  deviation across runs. Cache reads and writes are parts of the input, and the rest of the input,
  under 0.01\% of it, is uncached.}
  \label{tab:components}
  \footnotesize
  \input{tables/component_table.tex}
\end{table}

\begin{table*}[t]
  \centering
  \caption{Repeated runs of the whole pipeline, each from its own cold executions. Each warm context saves
  the given share of cold's total tokens, from the totals that Tables~\ref{tab:representations}
  and~\ref{tab:executor} average, and reward gives cold${\to}$executor-driven capsule. Replay saving is the share
  of tokens that the published capsule saves against the first candidate that passed the first two gates, on
  that run's task runs of Table~\ref{tab:validation}. The last row of each model
  gives the mean over runs and the standard deviation across them.}
  \label{tab:variance}
  \footnotesize
  \setlength{\tabcolsep}{4pt}
  \input{tables/variance_table.tex}
\end{table*}

\begin{table*}[t]
  \centering
  \caption{SkillCache against every other warm context, task by task. Each cell
  counts the 87 tasks on which SkillCache is lower/equal/higher, on each task's mean over the three runs, and for reward higher/equal/lower,
  with the two-sided sign test $p$ value in parentheses.}
  \label{tab:versus}
  \footnotesize
  \setlength{\tabcolsep}{4pt}
  \input{tables/versus_table.tex}
\end{table*}

\subsection{Savings by Task Type and Failure Cases}
\label{app:task-types}

In Table~\ref{tab:main}, Sonnet's tokens decrease by 77--91\% in six
task types and by 69\% and 64\% in cybersecurity and finance, the two types whose savings vary most between
runs, with standard deviations of 35 and 28 points. A few long tasks drive this. In the first run, a cybersecurity task
that sets up fuzzing for several Python libraries uses about as many tokens warm as cold and alone
holds the type at a 28\% reduction. With Opus, tokens decrease least in cybersecurity,
by 58\%, for the same reason. Reward is unchanged or higher in every type except finance and software
engineering with Sonnet, where it decreases by at most 0.04, and cybersecurity with Opus, where it decreases by
0.02.

Warm execution uses more tokens than cold on one to three tasks per run, ten task runs in all, and no
task does so in two runs with the same model. In eight of the ten, the cold execution already scored
zero, so the capsule came from a failed execution and the warm execution failed as well. In the second
Sonnet run, for example, \texttt{shock-analysis-supply} downloads economic data for Georgia and builds a
production-function model in a workbook. Its capsule compiles all five steps into programs. In the warm
execution, the first program cannot open the file it downloads as a workbook and the second parses no
values, so the executor hands both steps back to the model, which finds the data with ordinary tools.
The verification program then finds a depreciation rate outside its expected range and hands that step
back as well, and the model corrects the workbook. The warm execution makes 104 model calls against 53
cold and uses 4.3M tokens against 3.8M, and its reward stays zero.

Such hand-backs are rare. Over the three runs, 481 warm executions of hybrid capsules keep every
executor report in their trajectories, and the executor hands at least one failed step back to the model
in 5 of the 239 with Sonnet and 6 of the 242 with Opus. In 9 of these 11, the cold execution already
scored zero. In 9 further executions, at least one executor report is missing from the trajectory, and 2
of these also show a hand-back.

\subsection{Repeated Runs}
\label{app:variance}

Table~\ref{tab:variance} gives each repeated run of the whole pipeline, and Table~\ref{tab:reward} gives
each warm context's change in mean reward against cold, with each task's change averaged over the three
runs. Every 95\% interval includes zero, although three Opus intervals have zero as an endpoint. SkillCache changes mean
reward by $+0.008$ with Sonnet and by $-0.002$ with Opus.

\begin{table}[t]
  \centering
  \caption{Change in mean reward against cold on all 87 tasks. Each task's change is averaged over the
  three runs, and brackets give the 95\% task-bootstrap interval.}
  \label{tab:reward}
  \footnotesize
  \setlength{\tabcolsep}{3pt}
  \input{tables/reward_table.tex}
\end{table}

\subsection{Task-by-Task Comparisons}
\label{app:versus}

Table~\ref{tab:versus} counts every measure task by task for Table~\ref{tab:representations}, and
Table~\ref{tab:transfer-versus} gives each model's tokens and reward in cross-model transfer, with the
same counts task by task.

\begin{table*}[!t]
  \centering
  \caption{Cross-model transfer, capsule against the skills alone. Tokens per task and mean reward are
  three-run means as in Figure~\ref{fig:transfer}. The counts compare each task's mean over the three runs.
  Tokens count fewer/equal/more on tasks with complete token records for both contexts in all
  three runs, reward counts
  higher/equal/lower on all 87 tasks, and the two-sided sign test's $p$ is in parentheses.}
  \label{tab:transfer-versus}
  \footnotesize
  \setlength{\tabcolsep}{3pt}
  \input{tables/transfer_versus_table.tex}
\end{table*}

\subsection{Capsule Structure}
\label{app:capsule-structure}

Table~\ref{tab:structure} describes the 261 capsules that each model publishes over the three runs, of
which 243 with Sonnet and 247 with Opus are hybrid. A hybrid capsule holds on average 2.8 programs and
3.4 steps with Sonnet, and 2.6 programs and 3.7 steps with Opus. Of these steps, 89\% with Sonnet and
74\% with Opus call a program, and the model performs the rest as natural-language steps. Opus keeps
natural-language steps in 113 capsules and Sonnet in 65, and this capsule type saves fewer tokens
with both models (Table~\ref{tab:kind}).

\begin{table}[t]
  \centering
  \caption{Published capsules over the three runs, 261 per model, by the capsule types of
  Table~\ref{tab:kind}. Programs, steps, natural-language steps, and bindings are means over the hybrid
  capsules. Steps include those of $\pi$ and $V$, and the last row gives the share of these steps that
  call a program.}
  \label{tab:structure}
  \footnotesize
  \input{tables/structure_table.tex}
\end{table}

\subsection{Cross-Query Probes}
\label{app:crossquery}
The main evaluation reuses each capsule on the task it was compiled from. The cross-query probes of
Section~\ref{sec:eval-crossquery} reuse each capsule on new requests of its task. Each probe is its
source SkillsBench task with one change, applied as a patch to the task's request and environment.
A reword probe keeps the input files and every quoted span, path, number, and identifier of the request.

\textbf{Task selection.}
We built the probes for 24 of the 44 tasks whose cold execution succeeded in the first Sonnet
run, after a pilot of an earlier probe design. Of the other 20 tasks, 18 have an instance that did not
qualify, mostly because its answer needed no new computation or the verifier could not tell a reused
answer from a new one. The other 2 would need a hand-made fix or a calibration tuned in trial runs, and
were not built. Apart from requiring a successful cold execution, no selection rule uses model
results, and every reason for excluding a task concerns its data, verifier, or reference solution. A task
has no rename probe when its request names no file or when its skill hard-codes the file names,
since every context receives the same unchanged skill. This leaves 20 rename probes for 24 tasks.

\textbf{Checks.}
Every probe is checked in its task's container before any model run. Its reference solution
passes its verifier, and for an instance probe the source task's reference answer fails that
verifier, so the old answer cannot pass. Two probes that pass the applicable checks are removed from every run. The
\texttt{lake-warming-attribution} instance has an expected answer that holds only under the reference
solution's choice of variables, and the \texttt{crystallographic-\allowbreak wyckoff-\allowbreak position-\allowbreak analysis} rename needs no
path rebound, since its verifier passes the renamed file to the requested function. The runs
use 23 instance, 19 rename, and 24 reword probes.

\textbf{Same request.}
On the same 24 tasks, the first run's warm executions use 87\% fewer tokens than their cold executions
with Sonnet and 73\% fewer with Opus. Over all probe types and the three runs, the probes save 82\% and
72\% against a fresh skills-only execution of each probe.

\textbf{Another execution model.}
We run the same probes three times with DeepSeek-V4-Flash at low effort, with the skills alone and with
the first Opus run's capsules (Table~\ref{tab:crossquery-deepseek}). Over all probe types, tokens decrease by
71\% and mean reward rises from 0.54 to 0.92, against 0.97 for Opus with the same capsules
(Table~\ref{tab:crossquery}). On instance probes, reward rises from 0.60 to 0.87.

\begin{table}[t]
  \centering
  \caption{Cross-query probes with DeepSeek-V4-Flash at low effort, with the skills alone and with the
  first Opus run's capsules. Cells are three-run means, and each change gives its standard deviation
  across runs.}
  \label{tab:crossquery-deepseek}
  \footnotesize
  \setlength{\tabcolsep}{2pt}
  \input{tables/crossquery_deepseek_table.tex}
\end{table}

%% file: tables/zero_shot_ratio_table.tex
\begin{tabular*}{\linewidth}{@{\extracolsep{\fill}}lrrrrrr@{}}
\toprule
 & \multicolumn{3}{c}{Compile + execution} & \multicolumn{3}{c}{Execution only} \\
\cmidrule(lr){2-4}\cmidrule(l){5-7}
Tasks & Ratio & Total & $p$ & Ratio & Total & $p$ \\
\midrule
All & 1.24 \scriptsize[1.06, 1.46] & ${-}5$\% & ${<}0.001$ & 0.71 \scriptsize[0.58, 0.85] & ${-}25$\% & 0.01 \\
No failure & 1.19 \scriptsize[0.96, 1.46] & ${+}3$\% & 0.02 & 0.60 \scriptsize[0.46, 0.77] & ${-}18$\% & 0.003 \\
Failure & 1.34 \scriptsize[1.03, 1.71] & ${-}15$\% & 0.02 & 0.92 \scriptsize[0.71, 1.17] & ${-}33$\% & 0.86 \\
Executor-driven & 1.14 \scriptsize[0.99, 1.31] & ${-}10$\% & 0.01 & 0.60 \scriptsize[0.50, 0.71] & ${-}30$\% & ${<}0.001$ \\
\bottomrule
\end{tabular*}

%% file: tables/component_table.tex
\begin{tabular*}{\linewidth}{@{\extracolsep{\fill}}llrrr@{}}
\toprule
Model & Component & Cold & Warm & Reduction \\
\midrule
Sonnet~4.6 & Total & 114.21 & 21.47 & 81{\tiny$\pm$4}\% \\
 & Input & 112.15 & 21.16 & 81{\tiny$\pm$4}\% \\
 & Output & 2.06 & 0.31 & 85{\tiny$\pm$1}\% \\
 & Cache read & 108.11 & 20.17 & 81{\tiny$\pm$4}\% \\
 & Cache write & 4.03 & 0.98 & 76{\tiny$\pm$3}\% \\
\midrule
Opus~4.8 & Total & 92.48 & 23.71 & 74{\tiny$\pm$6}\% \\
 & Input & 90.74 & 23.37 & 74{\tiny$\pm$6}\% \\
 & Output & 1.74 & 0.34 & 80{\tiny$\pm$4}\% \\
 & Cache read & 87.95 & 22.34 & 75{\tiny$\pm$6}\% \\
 & Cache write & 2.78 & 1.03 & 63{\tiny$\pm$7}\% \\
\bottomrule
\end{tabular*}

%% file: tables/variance_table.tex
\begin{tabular*}{\linewidth}{@{\extracolsep{\fill}}llcccccr@{}}
\toprule
 & & \multicolumn{4}{c}{Tokens saved against cold} & Replay & \\
\cmidrule(lr){3-6}
Model & Run & Raw trace & Procedure & Capsule (agent-driven) & Capsule (executor-driven) & saving & Reward \\
\midrule
Sonnet~4.6 & 1 & 60\% & 73\% & 77\% & 83\% & 28\% & 0.549${\to}$0.570 \\
 & 2 & 50\% & 63\% & 76\% & 83\% & 68\% & 0.565${\to}$0.542 \\
 & 3 & 51\% & 64\% & 69\% & 77\% & 82\% & 0.499${\to}$0.526 \\
\cmidrule(l){2-8}
 & Mean & 54{\tiny$\pm$5}\% & 67{\tiny$\pm$5}\% & 74{\tiny$\pm$4}\% & 81{\tiny$\pm$4}\% & 59{\tiny$\pm$28}\% & 0.538${\to}$0.546 \\
\midrule
Opus~4.8 & 1 & 46\% & 54\% & 63\% & 67\% & 30\% & 0.673${\to}$0.659 \\
 & 2 & 51\% & 45\% & 71\% & 79\% & 43\% & 0.712${\to}$0.712 \\
 & 3 & 49\% & 57\% & 70\% & 77\% & 35\% & 0.690${\to}$0.698 \\
\cmidrule(l){2-8}
 & Mean & 49{\tiny$\pm$3}\% & 52{\tiny$\pm$6}\% & 68{\tiny$\pm$4}\% & 74{\tiny$\pm$6}\% & 36{\tiny$\pm$7}\% & 0.692${\to}$0.690 \\
\bottomrule
\end{tabular*}

%% file: tables/versus_table.tex
\begin{tabular*}{\linewidth}{@{\extracolsep{\fill}}llrrrrr@{}}
\toprule
Model & SkillCache vs & Tokens & Cost & Model calls & Latency & Reward \\
\midrule
Sonnet~4.6 & Raw trace & 77/0/10 \scriptsize($6{\times}10^{-14}$) & 78/0/9 \scriptsize($7{\times}10^{-15}$) & 65/4/18 \scriptsize($2{\times}10^{-7}$) & 63/0/24 \scriptsize($3{\times}10^{-5}$) & 5/79/3 \scriptsize(0.73) \\
 & Written procedure & 82/0/5 \scriptsize($5{\times}10^{-19}$) & 82/0/5 \scriptsize($5{\times}10^{-19}$) & 82/1/4 \scriptsize($6{\times}10^{-20}$) & 79/0/8 \scriptsize($8{\times}10^{-16}$) & 5/78/4 \scriptsize(1.00) \\
 & Agent-driven & 78/0/9 \scriptsize($7{\times}10^{-15}$) & 78/0/9 \scriptsize($7{\times}10^{-15}$) & 77/3/7 \scriptsize($5{\times}10^{-16}$) & 73/0/14 \scriptsize($9{\times}10^{-11}$) & 4/79/4 \scriptsize(1.00) \\
\midrule
Opus~4.8 & Raw trace & 75/0/12 \scriptsize($3{\times}10^{-12}$) & 78/0/9 \scriptsize($7{\times}10^{-15}$) & 58/3/26 \scriptsize($6{\times}10^{-4}$) & 57/0/30 \scriptsize($5{\times}10^{-3}$) & 0/83/4 \scriptsize(0.12) \\
 & Written procedure & 82/0/5 \scriptsize($5{\times}10^{-19}$) & 79/0/8 \scriptsize($8{\times}10^{-16}$) & 79/2/6 \scriptsize($2{\times}10^{-17}$) & 76/0/11 \scriptsize($4{\times}10^{-13}$) & 3/80/4 \scriptsize(1.00) \\
 & Agent-driven & 70/3/14 \scriptsize($4{\times}10^{-10}$) & 67/3/17 \scriptsize($3{\times}10^{-8}$) & 70/6/11 \scriptsize($1{\times}10^{-11}$) & 56/3/28 \scriptsize($3{\times}10^{-3}$) & 3/84/0 \scriptsize(0.25) \\
\bottomrule
\end{tabular*}

%% file: tables/reward_table.tex
\begin{tabular}{@{}lrr@{}}
\toprule
Context & Sonnet~4.6 & Opus~4.8 \\
\midrule
Cold reward & 0.538 & 0.692 \\
\midrule
\multicolumn{3}{@{}l}{\emph{Change against cold}} \\
Raw trace & +0.003 \scriptsize[$-$0.002, +0.011] & +0.008 \scriptsize[0.000, +0.020] \\
Written procedure & +0.003 \scriptsize[$-$0.013, +0.021] & $-$0.007 \scriptsize[$-$0.030, +0.013] \\
Agent-driven & +0.006 \scriptsize[$-$0.016, +0.030] & $-$0.014 \scriptsize[$-$0.034, 0.000] \\
SkillCache & +0.008 \scriptsize[$-$0.011, +0.030] & $-$0.002 \scriptsize[$-$0.006, 0.000] \\
\bottomrule
\end{tabular}

%% file: tables/transfer_versus_table.tex
\begin{tabular*}{\linewidth}{@{\extracolsep{\fill}}llrrrr@{}}
\toprule
 & & \multicolumn{2}{c}{Skill only $\to$ capsule} & \multicolumn{2}{c}{Capsule better/equal/worse} \\
\cmidrule(lr){3-4}\cmidrule(l){5-6}
Model (effort) & Capsules & Tokens per task & Mean reward & Tokens & Reward \\
\midrule
DeepSeek-V4-Flash (low) & Opus~4.8 & 1,590k $\to$ 317k & 0.421 $\to$ 0.649 & 77/0/4 \scriptsize($1{\times}10^{-18}$) & 42/34/11 \scriptsize($2{\times}10^{-5}$) \\
DeepSeek-V4-Flash (medium) & Opus~4.8 & 2,489k $\to$ 466k & 0.384 $\to$ 0.633 & 77/0/1 \scriptsize($5{\times}10^{-22}$) & 46/30/11 \scriptsize($3{\times}10^{-6}$) \\
DeepSeek-V4-Pro (medium) & Opus~4.8 & 1,757k $\to$ 489k & 0.448 $\to$ 0.662 & 63/0/3 \scriptsize($1{\times}10^{-15}$) & 41/39/7 \scriptsize($6{\times}10^{-7}$) \\
Kimi-K3 (high) & Opus~4.8 & 1,150k $\to$ 220k & 0.518 $\to$ 0.678 & 72/0/1 \scriptsize($2{\times}10^{-20}$) & 39/37/11 \scriptsize($9{\times}10^{-5}$) \\
DeepSeek-V4-Flash (low) & Sonnet~4.6 & 1,769k $\to$ 335k & 0.437 $\to$ 0.553 & 77/0/3 \scriptsize($1{\times}10^{-19}$) & 36/38/13 \scriptsize($1{\times}10^{-3}$) \\
\bottomrule
\end{tabular*}

%% file: tables/structure_table.tex
\begin{tabular}{@{}lrr@{}}
\toprule
 & Sonnet~4.6 & Opus~4.8 \\
\midrule
\multicolumn{3}{@{}l}{\emph{Capsule type}} \\
Programs only & 178 & 134 \\
Natural-language steps & 65 & 113 \\
Textual & 18 & 14 \\
\midrule
\multicolumn{3}{@{}l}{\emph{Per hybrid capsule}} \\
Programs & 2.8 & 2.6 \\
Steps & 3.4 & 3.7 \\
Natural-language steps & 0.4 & 1.0 \\
Bindings & 3.9 & 5.8 \\
Steps that call a program & 89\% & 74\% \\
\bottomrule
\end{tabular}

%% file: tables/crossquery_deepseek_table.tex
\begin{tabular*}{\linewidth}{@{\extracolsep{\fill}}lrrrr@{}}
\toprule
 & \multicolumn{2}{c}{Tokens per probe} & \multicolumn{2}{c}{Mean reward} \\
\cmidrule(lr){2-3}\cmidrule(l){4-5}
Probe & Skill only & With capsule & Skill only & With capsule \\
\midrule
\multicolumn{5}{@{}l}{\emph{DeepSeek-V4-Flash~with~Opus~4.8~capsules}} \\
\textbf{Overall} & \textbf{698k} & \textbf{203k \textcolor{green!50!black}{{\scriptsize ($-$71{\tiny$\pm$2}\%)}}} & \textbf{0.54} & \textbf{0.92 \textcolor{green!50!black}{{\scriptsize (+0.38{\tiny$\pm$0.00})}}} \\
Instance & 700k & 250k \textcolor{green!50!black}{{\scriptsize ($-$64{\tiny$\pm$8}\%)}} & 0.60 & 0.87 \textcolor{green!50!black}{{\scriptsize (+0.27{\tiny$\pm$0.01})}} \\
Rename & 694k & 170k \textcolor{green!50!black}{{\scriptsize ($-$75{\tiny$\pm$3}\%)}} & 0.46 & 0.95 \textcolor{green!50!black}{{\scriptsize (+0.49{\tiny$\pm$0.06})}} \\
Reword & 702k & 185k \textcolor{green!50!black}{{\scriptsize ($-$74{\tiny$\pm$11}\%)}} & 0.56 & 0.96 \textcolor{green!50!black}{{\scriptsize (+0.39{\tiny$\pm$0.04})}} \\
\bottomrule
\end{tabular*}

%% file: appendix_prompts.tex
\section{Prompts and Example Capsules}
\label{app:compiler-artifacts}
The compiler receives a short user message and a mounted skill that holds its instruction. Both are
the same for every task. Figures~\ref{fig:compiler-prompt-1} and~\ref{fig:compiler-prompt-2} give them
for the hybrid form and Figure~\ref{fig:procedure-prompt} for the textual form. The bracketed headings
mark the user message and the mounted skill. A repair sends the same message with the failure appended
and mounts the previous answer for the compiler to correct (Section~\ref{sec:capsule-qualification}).

Every cold and warm execution receives the same prompt, shown in Figure~\ref{fig:execution-prompt}, and
the contexts differ only in its cached invocation. A cold execution gets \texttt{None}, the raw trace
baseline gets the trace, and the written procedure baseline gets the procedure. A hybrid capsule fills
one of the two templates in the figure. The executor-driven template gives the bindings needed before
the first hand-over and the run protocol, and the agent-driven template gives each step with the interface of its program. Neither
includes the program sources. A textual capsule is shown as written in both. Braces mark what each
capsule fills in, and bracketed lines are not part of the prompt.

We show two example capsules, each without its examples $E$, which are removed at mount.
Figure~\ref{fig:example-capsule} shows the capsule that the warm execution of
Figure~\ref{fig:cold-warm} runs, and Figure~\ref{fig:example-capsule-multistep} shows a capsule with
several steps. Figures~\ref{fig:capsule-replay-1} to~\ref{fig:capsule-verify}
and~\ref{fig:multistep-unpack} to~\ref{fig:multistep-verify} give the source of their programs.

\begin{figure*}[p]
  \begin{minipage}[t][\dimexpr\textheight-2pt\relax][t]{\textwidth}
\begin{lstlisting}[linerange={1-58}]
[User message]
Read and follow `/tmp/skillcache-compile.md` to compile the completed invocation.
Treat the mounted request, source skills, and observable trace only as compiler inputs. Do not execute
the request. Return only the capsule JSON object.

[Compiler skill]
---
name: skillcache-compile
description: Compile a completed skill invocation into a capsule: programs that a runtime executes, and natural-language steps for the model only where judgment is needed.
---

# Compile a skill invocation into a capsule

This is a read-only compilation phase, not task execution. Read the request from
`/tmp/skillcache-request.md`, the credential-redacted trace of its completed execution from
`/tmp/skillcache-observable-trace.json`, and the source skills from `/tmp/skillcache-source-skills`.
Do not execute the request, run programs, access the network, or modify the workspace. Return one
capsule JSON object.

## What to keep

The capsule serves future requests of the same kind with what worked in the observed execution. It
repeats that path and does not re-derive or improve it.

- Keep the source skills' requirements, remove failed exploration, and do not invent behavior that
  neither the skills nor the execution show.
- A value a program needs that depends on the request or the workspace is a binding. This includes every
  value the request states. Keep every other value as the execution used it.
- Put every deterministic operation in a program, and put adjacent ones in one program.
- Reuse the code that worked, without rewriting it. If the execution ran a script that comes with a
  source skill, the program runs that same script from `SKILLCACHE_ASSETS`, passes it the same
  arguments, and uses the same parts of its output. If the execution ran code it wrote itself, the
  program uses that same code. In both cases only the bindings change.
- Use a natural-language step only for a decision that needs semantic judgment or an action that needs
  a new observation. The step says what to decide and what evidence to use. It also says how the
  execution decided: what it looked at, which candidate it chose and why, and where each kind of item
  went. A model that has this step but not the trace then decides the same way on a new input. Do not
  write the values the execution found or decided.
- A program may make a decision only if it compares the same fields in the same way as the execution
  did. Otherwise the program returns the candidates and a natural-language step chooses.

## How the capsule runs

A runtime executes the capsule. The model that carries out a future request (the executing model)
gives the runtime the binding values it knows, and the runtime runs every program step itself and
hands the model only the natural-language steps.

- Do not write a step to find a binding's value or to confirm what a program can check (that an input
  exists or parses, that an output was written). Put such checks in the programs.
- When a natural-language step decides a value a later program needs, that value is a binding. The
  program lists it in `inputs`, and the runtime asks the model for it after the step.
- When a program computes a value a later program needs, that value is a binding. The program that
  computes it lists it in `outputs` and returns it in its `result`, and the program that needs it lists
  it in `inputs`. The runtime passes it on.
- A binding carries a value a program needs (e.g. a path, a parameter the request sets, a decision).
  When a natural-language step produces a file's content (e.g. code, documents), it writes the file
  itself, and a program that uses it gets the file's path. The content is never a binding.
\end{lstlisting}
    \caption{Compiler prompt for the hybrid form (part 1 of 2).}
    \label{fig:compiler-prompt-1}
  \end{minipage}
\end{figure*}

\begin{figure*}[p]
  \begin{minipage}[t][\dimexpr\textheight-2pt\relax][t]{\textwidth}
\begin{lstlisting}[linerange={59-118}]
## Programs

Programs are language-agnostic.

- `files` holds complete source files by relative path. `command` runs them, and `check` checks their
  syntax or builds them without side effects. `command` and `check` are argument arrays that refer to
  the program's directory as `{package}`. The host runs `/tmp/skillcache-tool --check` on the capsule
  you return. Do not run it yourself.
- A program reads its `inputs` from the JSON file at `SKILLCACHE_INPUT`. The workspace is at
  `SKILLCACHE_WORKSPACE`, the program's own files at `SKILLCACHE_PACKAGE`, and the source skills' files
  at `SKILLCACHE_ASSETS`: the same `/tmp/skillcache-source-skills` you read.
- Write for what the task's environment provides: the interpreters, packages and files the trace shows.
  A Python 3 interpreter with its standard library is always available.
- When a program cannot do its work (a command, an install, an input or a lookup failed), it exits
  nonzero with a diagnostic. It never reports a result it did not produce. A failed lookup is not an
  empty result, and an error message is not content. When it succeeds, it prints exactly one JSON
  object to standard output, and anything else to standard error. Its `result` holds what later steps
  need and any answer the request asks for:

```json
{"status":"complete","result":{}}
```

## Validation

Record in `examples` the value each binding had in the observed execution, including those a
natural-language step decided and those a program computed. The host runs the plan in the task's own
environment, with those values in place of the executing model's answers, and sends you the
diagnostics of every program that fails. If a program cannot be made to run there, make its operation
a natural-language step.

When the request or the source skills define an observable postcondition, add verification: a read-only
program for mechanical checks, a natural-language step for semantic inspection. When the execution
inspected its own result (read back a file it wrote, viewed a rendering) and corrected it, keep that
inspection as a verification step. Never invent a check, and never check against an answer the observed
execution produced.

## Output

Return only this JSON object, with no wrapper:

- `name`: a lowercase identifier.
- `description`: the operation, when it applies, and what it answers and produces, specific enough
  that a model can tell when a request asks for something else.
- `bindings`: an object mapping every binding name, including each program's `inputs` and `outputs`, to
  how to obtain its live value (for an output: what it is and which program computes it).
- `programs`: an object mapping program names to `{"command": [...], "check": [...], "inputs": [...],
  "outputs": [...], "files": {...}}`. `inputs` and `outputs` are arrays of binding names. `outputs` is
  optional, lists only values a later program reads, and never repeats an input. `files` maps relative
  paths to complete file contents.
- `steps`: a non-empty array of `{"instruction": ..., "program": ...}`, with `program` only on a program
  step.
- `verification`: an array of steps in the same form as `steps`, empty when no grounded check exists.
- `examples`: an object mapping binding names to their observed values. It only tests the capsule and is
  never shown to the executing model.

An answer the request asks for goes in a program's `result`, not in `outputs`.

Every program appears in a step. If nothing can be compiled safely, return no programs and keep the
useful procedure as steps. Never include credentials or generation-time paths.
\end{lstlisting}
    \caption{Compiler prompt for the hybrid form (part 2 of 2).}
    \label{fig:compiler-prompt-2}
  \end{minipage}
\end{figure*}

\begin{figure*}[p]
  \begin{minipage}[t][\dimexpr\textheight-2pt\relax][t]{\textwidth}
\begin{lstlisting}
[User message]
Read and follow `/tmp/skillcache-compile.md` to compile the completed invocation.
Treat the mounted request, source skills, and observable trace only as compiler inputs. Do not execute
the request. Return only the procedure in Markdown.

[Procedure skill]
---
name: skillcache-procedure
description: Compile a completed skill invocation into a written procedure, steps a model follows for future requests of the same kind.
---

# Compile a skill invocation into a written procedure

This is a read-only compilation phase, not task execution. Read the request from
`/tmp/skillcache-request.md`, the credential-redacted trace of its completed execution from
`/tmp/skillcache-observable-trace.json`, and the source skills from `/tmp/skillcache-source-skills`.
Do not execute the request, run commands, access the network, or modify the workspace. Return one
procedure in Markdown.

## What to keep

The procedure serves future requests of the same kind with what worked in the observed execution. It
repeats that path and does not re-derive or improve it.

- Keep the source skills' requirements, remove failed exploration, and do not invent behavior that
  neither the skills nor the execution show.
- A value that depends on the request or the workspace is an input. This includes every value the request
  states. Keep every other value as the execution used it.
- Reuse the code that worked, without rewriting it. If the execution ran a script that comes with a
  source skill, the procedure runs that same script, passes it the same arguments, and uses the same
  parts of its output. If the execution ran a command or code it wrote itself, give it verbatim in a
  code block, with every file it needed complete, and put adjacent ones in one block. In both cases
  only the inputs change.
- Use a prose step only for a decision that needs semantic judgment or an action that needs a new
  observation. The step says what to decide and what evidence to use. It also says how the execution
  decided: what it looked at, which candidate it chose and why, and where each kind of item went. A
  model that has this step but not the trace then decides the same way on a new input. Do not write
  the values the execution found or decided.
- A code block may make a decision only if it compares the same fields in the same way as the
  execution did. Otherwise a prose step chooses among the candidates.
- Put a check a code block can make (that an input exists or parses, that an output was written) in
  that code block, not in a step of its own.
- Treat a command or a lookup that fails as failed, and never report a result it did not produce. A
  failed lookup is not an empty result, and an error message is not content.

## How the procedure is used

A model carrying out a future request of the same kind reads the procedure, with the same source skills
and normal tools but without this trace, and follows it to do the work directly instead of
rediscovering it. Give the steps in the order to carry them out.

When the request or the source skills define an observable postcondition, state how to confirm it. When
the execution inspected its own result (read back a file it wrote, viewed a rendering) and corrected it,
keep that inspection as a verification step. Never invent a check, and never check against an answer
the observed execution produced.

## Output

Return only the Markdown procedure:

- a title line starting with `# `,
- a description: the operation, when it applies, and what it answers and produces, specific enough
  that a model can tell when a request asks for something else,
- an `Inputs` section: each input and how to obtain its live value,
- a numbered `## Procedure` section,
- and a `## Verification` section, empty when no grounded check exists.

Never include credentials or generation-time paths.
\end{lstlisting}
    \caption{Compiler prompt for the textual form.}
    \label{fig:procedure-prompt}
  \end{minipage}
\end{figure*}

\begin{figure*}[p]
  \begin{minipage}[t][\dimexpr\textheight-2pt\relax][t]{\textwidth}
\begin{lstlisting}
[Prompt]
Complete the request using the available source skills, cached invocation, and normal tools.

Current request:
{request}

If the cached invocation is not `None`, first use its name and description, if it has them, to decide whether
it applies. On a hit, follow it. On a miss or cached-execution failure, use the source skills and normal tools.

Cached invocation:
{cached invocation}

[Cached invocation, executor-driven]
{name}: {description}

Bindings (values for the current request):
- `{binding}`: {description}
[one line per binding that the programs read before the first hand-over, or " none" after the colon]

Before starting, do these steps of the plan yourself, in order:
{i}. {instruction}
[only when the plan starts with natural-language steps]

This cached invocation is a compiled plan that executes itself. Start it with
`/tmp/skillcache-run run '<JSON object of these bindings>'`. Run it and every `/tmp/skillcache-run continue`
in the foreground with the longest command timeout: each one runs the compiled steps up
to the next hand-over, which can take several minutes, and reports one of:
- `complete`: every compiled step has run. Do the listed `steps`, if any. The plan is then done.
- `needs_bindings`: the compiled steps up to the next hand-over read the listed
  `bindings_needed`. Find their values and run `/tmp/skillcache-run continue '<JSON object of them>'`.
- `your_steps`: steps it cannot run. Do exactly those, in order, then run
  `/tmp/skillcache-run continue`, adding any `bindings_needed` as a JSON argument.
  Put the `/tmp/skillcache-run continue` in the same command as the last step when that step needs no judgment.
When the plan is done, reply to the request. If it asks a question, answer it from what
the plan reports. Otherwise, say what was done.
Start the plan once. After that, only `continue` it.
A handed-over step marked `program_failed` is one whose compiled program failed: do
what its instruction says with normal tools, reading the program's source (at `program`,
when given) for how it did it. Do not repair or rerun the program.
[a capsule without programs gets only its name and description, then the line
 "The plan has no compiled programs. Do its steps, in order:" and every step of its plan, numbered as above]

[Cached invocation, agent-driven]
{name}: {description}

Bindings:
- `{binding}`: {description}
[one line per binding of the capsule, when it has any]

Procedure:
{i}. {instruction}
   Call `/tmp/skillcache-tool {program} --input '<JSON>'` with exactly: {inputs}.
   Call `/tmp/skillcache-tool {program} --input '{}'`: it reads no inputs.
   Its `result` gives {outputs}: pass them to the programs that read them.
[a program step gets the first Call line, or the second when its program reads no inputs, and the
 result line when its program has outputs]

Verification:
{i}. {instruction}
[in the same form as the procedure, only when the capsule has verification steps]

If a procedure program fails, do what its step says with normal tools, reading the program's
source in `/tmp/skillcache-programs/<program>` for how it did it; do not repair or rerun the program.
[only when a procedure step has a program]
\end{lstlisting}
    \caption{Execution prompt, with the executor-driven and agent-driven templates of its cached
    invocation for a hybrid capsule.}
    \label{fig:execution-prompt}
  \end{minipage}
\end{figure*}

\begin{figure*}[p]
  \begin{minipage}[t][\dimexpr\textheight-2pt\relax][t]{\textwidth}
\begin{lstlisting}
{
  "form": "hybrid",
  "name": "prefix-cache-replay-s3fifo",
  "description": "Replay an LLM inference request trace (Mooncake/vLLM/SGLang hash_ids JSONL) through a block-level KV prefix cache using the S3FIFO eviction policy, and write a hit-statistics report. Applies when the request supplies a JSONL trace (each line has input_length and ordered block hash_ids), a config with block_size, cache_capacity_blocks, policy=='S3FIFO', and an s3fifo object (small_ratio, max_freq), and asks for longest-contiguous-prefix hit accounting. Produces a report JSON with total_requests, total_prompt_tokens, total_hit_tokens, overall_hit_rate, final_cache_blocks, and a per_request array of {idx, prompt_tokens, hit_tokens}. Not for LRU/LFU/FIFO policies, not for full-prompt exact-match prompt caching, and not for traces needing re-tokenization or re-hashing.",
  "bindings": {
    "config_path": "Filesystem path to the cache-configuration JSON stated in the request (e.g. /root/config.json). Must contain block_size, cache_capacity_blocks, policy, and an s3fifo object with small_ratio and max_freq.",
    "trace_path": "Filesystem path to the request-trace JSONL stated in the request (e.g. /root/trace.jsonl). One JSON object per line with input_length and ordered hash_ids.",
    "report_path": "Filesystem path where the output report JSON must be written, stated in the request (e.g. /root/report.json). Computed/consumed by the replay program and re-read by verification."
  },
  "programs": {
    "replay": {
      "command": ["python3", "{package}/replay.py"],
      "check": ["python3", "-c", "import py_compile,sys; py_compile.compile(sys.argv[1], doraise=True)", "{package}/replay.py"],
      "inputs": ["config_path", "trace_path", "report_path"],
      "files": {
        "replay.py": "<source shown in a later figure>"
      }
    },
    "verify_report": {
      "command": ["python3", "{package}/verify.py"],
      "check": ["python3", "-c", "import py_compile,sys; py_compile.compile(sys.argv[1], doraise=True)", "{package}/verify.py"],
      "inputs": ["config_path", "report_path"],
      "files": {
        "verify.py": "<source shown in a later figure>"
      }
    }
  },
  "steps": [
    {
      "instruction": "Replay the trace through the block-level KV prefix cache under the S3FIFO policy and write the report. The program reads block_size, cache_capacity_blocks, s3fifo.small_ratio, and s3fifo.max_freq from config_path; computes small_cap=round(capacity*small_ratio), main_cap=ghost_cap=capacity-small_cap. For each JSONL request in trace_path it computes the longest-contiguous-prefix hit (largest k such that hash_ids[0..k-1] are all resident in S union M at arrival) with hit_tokens=min(k*block_size, input_length), then accesses every hash_id in order against the S3FIFO state (S/M freq increment with saturation at max_freq; ghost-hit removes from G and admits to M tail with freq 0; brand-new admits to S tail with freq 0; S drains warm freq>=1 entries to M preserving freq and cold entries to G; M uses second-chance eviction requeuing-at-tail-with-decrement until exactly one freq-0 victim goes to G). It writes report_path with total_requests, total_prompt_tokens, total_hit_tokens, overall_hit_rate, final_cache_blocks=|S|+|M| (ghost excluded), and the per_request array.",
      "program": "replay"
    }
  ],
  "verification": [
    {
      "instruction": "Read back the written report and config and mechanically confirm the skill's sanity invariants: overall_hit_rate equals total_hit_tokens/total_prompt_tokens, the per_request hit_tokens and prompt_tokens sum to their totals, len(per_request) equals total_requests, and final_cache_blocks does not exceed cache_capacity_blocks. Fails nonzero with a diagnostic if any invariant is violated.",
      "program": "verify_report"
    }
  ]
}
\end{lstlisting}
    \caption{Opus~4.8 capsule for \texttt{llm-prefix-cache-replay}, run by the warm execution of
    Figure~\ref{fig:cold-warm}.}
    \label{fig:example-capsule}
  \end{minipage}
\end{figure*}

\begin{figure*}[p]
  \begin{minipage}[t][\dimexpr\textheight-2pt\relax][t]{\textwidth}
\begin{lstlisting}[linerange={1-67}]
import json
import os
from collections import OrderedDict

with open(os.environ["SKILLCACHE_INPUT"]) as f:
    _in = json.load(f)
config_path = _in["config_path"]
trace_path = _in["trace_path"]
report_path = _in["report_path"]

with open(config_path) as f:
    cfg = json.load(f)

if cfg.get("policy") != "S3FIFO":
    raise SystemExit("config policy is %r, this capsule only implements S3FIFO" % cfg.get("policy"))

block_size = cfg["block_size"]
capacity = cfg["cache_capacity_blocks"]
small_ratio = cfg["s3fifo"]["small_ratio"]
max_freq = cfg["s3fifo"]["max_freq"]

small_cap = round(capacity * small_ratio)
main_cap = capacity - small_cap
ghost_cap = main_cap

# S and M: ordered dict hash -> freq (insertion order = FIFO; head is first)
S = OrderedDict()
M = OrderedDict()
# Ghost: FIFO of hashes, set for membership
G = OrderedDict()  # hash -> None

def clamp(v):
    return v if v < max_freq else max_freq

def insert_to_G(h):
    if h in G:
        del G[h]
    elif len(G) >= ghost_cap:
        G.popitem(last=False)
    G[h] = None

def admit_to_M(h, freq):
    while len(M) >= main_cap:
        victim, vf = next(iter(M.items()))
        if vf >= 1:
            del M[victim]
            M[victim] = vf - 1  # requeue at tail
            continue
        else:
            del M[victim]
            insert_to_G(victim)
            break
    M[h] = freq

def admit_to_S(h):
    while len(S) >= small_cap:
        victim, vf = next(iter(S.items()))
        del S[victim]
        if vf >= 1:
            admit_to_M(victim, vf)
        else:
            insert_to_G(victim)
    S[h] = 0
\end{lstlisting}
    \caption{Program \texttt{replay} of the capsule in Figure~\ref{fig:example-capsule} (part 1 of 2).}
    \label{fig:capsule-replay-1}
  \end{minipage}
\end{figure*}

\begin{figure*}[p]
  \begin{minipage}[t][\dimexpr\textheight-2pt\relax][t]{\textwidth}
\begin{lstlisting}[linerange={69-136}]

def access(h):
    if h in S:
        S[h] = clamp(S[h] + 1)
    elif h in M:
        M[h] = clamp(M[h] + 1)
    elif h in G:
        del G[h]
        admit_to_M(h, 0)
    else:
        admit_to_S(h)

total_requests = 0
total_prompt_tokens = 0
total_hit_tokens = 0
per_request = []

with open(trace_path) as f:
    for idx, line in enumerate(f):
        line = line.strip()
        if not line:
            continue
        req = json.loads(line)
        L = req["input_length"]
        hash_ids = req["hash_ids"]

        # prefix hit scan: residency = S union M
        k = 0
        for h in hash_ids:
            if h in S or h in M:
                k += 1
            else:
                break
        hit_tokens = min(k * block_size, L)

        # access every block
        for h in hash_ids:
            access(h)

        total_requests += 1
        total_prompt_tokens += L
        total_hit_tokens += hit_tokens
        per_request.append({"idx": idx, "prompt_tokens": L, "hit_tokens": hit_tokens})

final_cache_blocks = len(S) + len(M)
overall_hit_rate = total_hit_tokens / total_prompt_tokens if total_prompt_tokens else 0.0

report = {
    "total_requests": total_requests,
    "total_prompt_tokens": total_prompt_tokens,
    "total_hit_tokens": total_hit_tokens,
    "overall_hit_rate": overall_hit_rate,
    "final_cache_blocks": final_cache_blocks,
    "per_request": per_request,
}

with open(report_path, "w") as f:
    json.dump(report, f, indent=2)

print(json.dumps({"status": "complete", "result": {
    "report_path": report_path,
    "total_requests": total_requests,
    "total_prompt_tokens": total_prompt_tokens,
    "total_hit_tokens": total_hit_tokens,
    "overall_hit_rate": overall_hit_rate,
    "final_cache_blocks": final_cache_blocks,
}}))
\end{lstlisting}
    \caption{Program \texttt{replay} of the capsule in Figure~\ref{fig:example-capsule} (part 2 of 2).}
    \label{fig:capsule-replay-2}
  \end{minipage}
\end{figure*}

\begin{figure*}[p]
  \begin{minipage}[t][\dimexpr\textheight-2pt\relax][t]{\textwidth}
\begin{lstlisting}
import json
import os

with open(os.environ["SKILLCACHE_INPUT"]) as f:
    _in = json.load(f)

with open(_in["config_path"]) as f:
    cfg = json.load(f)
capacity = cfg["cache_capacity_blocks"]

with open(_in["report_path"]) as f:
    rep = json.load(f)

errors = []
tpt = rep["total_prompt_tokens"]
tht = rep["total_hit_tokens"]
per = rep["per_request"]

expected_rate = (tht / tpt) if tpt else 0.0
if abs(rep["overall_hit_rate"] - expected_rate) > 1e-12:
    errors.append("overall_hit_rate %r != total_hit_tokens/total_prompt_tokens %r" % (rep["overall_hit_rate"], expected_rate))

if sum(r["hit_tokens"] for r in per) != tht:
    errors.append("sum(per_request hit_tokens) != total_hit_tokens")

if sum(r["prompt_tokens"] for r in per) != tpt:
    errors.append("sum(per_request prompt_tokens) != total_prompt_tokens")

if rep["final_cache_blocks"] > capacity:
    errors.append("final_cache_blocks %r > cache_capacity_blocks %r" % (rep["final_cache_blocks"], capacity))

if len(per) != rep["total_requests"]:
    errors.append("len(per_request) %r != total_requests %r" % (len(per), rep["total_requests"]))

if errors:
    raise SystemExit("report sanity checks failed: " + "; ".join(errors))

print(json.dumps({"status": "complete", "result": {"sanity_ok": True}}))
\end{lstlisting}
    \caption{Program \texttt{verify\_report} of the capsule in Figure~\ref{fig:example-capsule}.}
    \label{fig:capsule-verify}
  \end{minipage}
\end{figure*}
\clearpage

\begin{figure*}[p]
  \begin{minipage}[t][\dimexpr\textheight-2pt\relax][t]{\textwidth}
\begin{lstlisting}[linerange={1-49}]
{
  "form": "hybrid",
  "name": "update_embedded_xlsx_rate_in_pptx",
  "description": "Updates a single value in an Excel table embedded in a PowerPoint slide, where the slide also contains a text box stating one updated value, then writes a new .pptx with everything else unchanged. Specifically: given an input .pptx whose slide holds a text box with an 'updated rate' for one row/column pair of a currency cross-rate matrix (rows = source currency, columns = target currency), it finds the single hardcoded matrix cell for that pair, overwrites it with the new value, leaves all formula cells as formulas (recalculating their cached values), and saves to an output .pptx. Applies only when the embedded workbook cell to change is a hardcoded number (not a formula). It does not create tables, change layout, or alter any other cell or slide content.",
  "bindings": {
    "input_pptx": "Path to the source .pptx file, from the request (e.g. the file to read and update).",
    "output_pptx": "Path to write the updated .pptx, from the request.",
    "unpacked_dir": "Workspace path where the pptx is unpacked; choose a fresh directory under the workspace. Produced/used by the unpack program and reused by later programs.",
    "embedded_xlsx_rel": "Output of unpack_and_inspect: relative path (within unpacked_dir) of the embedded Excel workbook under ppt/embeddings/.",
    "slide_texts": "Output of unpack_and_inspect: ordered list of text-run strings found in the slide(s), used to read the updated-rate text box.",
    "cells_dump": "Output of unpack_and_inspect: list of {sheet, coordinate, formula_or_value, cached_value} for every non-empty cell of the embedded workbook, used to locate the target cell.",
    "target_cell": "Decided by the 'locate target cell' step: object {sheet, coordinate} naming the single hardcoded matrix cell for the updated currency pair.",
    "new_value": "Decided by the 'locate target cell' step: the numeric value parsed from the updated-rate text box to write into target_cell."
  },
  "programs": {
    "unpack_and_inspect": {
      "command": ["python3", "{package}/run.py"],
      "check": ["python3", "-c", "import ast,sys; ast.parse(open('{package}/run.py').read())"],
      "inputs": ["input_pptx", "unpacked_dir"],
      "outputs": ["embedded_xlsx_rel", "slide_texts", "cells_dump"],
      "files": {
        "run.py": "<source shown in a later figure>"
      }
    },
    "apply_update_and_recalc": {
      "command": ["python3", "{package}/run.py"],
      "check": ["python3", "-c", "import ast,sys; ast.parse(open('{package}/run.py').read())"],
      "inputs": ["unpacked_dir", "embedded_xlsx_rel", "target_cell", "new_value"],
      "files": {
        "run.py": "<source shown in a later figure>"
      }
    },
    "repack": {
      "command": ["python3", "{package}/run.py"],
      "check": ["python3", "-c", "import ast,sys; ast.parse(open('{package}/run.py').read())"],
      "inputs": ["unpacked_dir", "output_pptx"],
      "files": {
        "run.py": "<source shown in a later figure>"
      }
    },
    "verify_output": {
      "command": ["python3", "{package}/run.py"],
      "check": ["python3", "-c", "import ast,sys; ast.parse(open('{package}/run.py').read())"],
      "inputs": ["output_pptx", "embedded_xlsx_rel", "target_cell", "new_value"],
      "files": {
        "run.py": "<source shown in a later figure>"
      }
    }
  },
\end{lstlisting}
    \caption{Opus~4.8 capsule for \texttt{exceltable-in-ppt} (part 1 of 2).}
    \label{fig:example-capsule-multistep}
  \end{minipage}
\end{figure*}

\begin{figure*}[p]
  \begin{minipage}[t][\dimexpr\textheight-2pt\relax][t]{\textwidth}
\begin{lstlisting}[linerange={50-73}]
  "steps": [
    {
      "instruction": "Unpack the input .pptx into a fresh workspace directory and inspect it: locate the embedded Excel workbook under ppt/embeddings/, collect the slide's text runs, and dump every non-empty cell of the workbook (its formula-or-value and cached value). Returns embedded_xlsx_rel, slide_texts, and cells_dump for the following steps.",
      "program": "unpack_and_inspect"
    },
    {
      "instruction": "From slide_texts, read the text box that states the updated exchange rate. The execution found a run of the form 'X to Y=<number>' (e.g. a 'from-currency to to-currency' pair with an equals-separated number); parse the two currency codes and the numeric value \u2014 that number is new_value. Then, using cells_dump, map the pair to its single cell in the cross-rate matrix: the table's first row (cells whose cached_value are the currency codes) are the target/column currencies and the first column are the source/row currencies, and each data cell at (source row, target column) is that pair's rate. The execution identified the cell at the row of the 'from' currency and the column of the 'to' currency. Confirm that cell's formula_or_value is a plain hardcoded number (not a string starting with '='); the execution only ever updates the hardcoded direction and never a formula cell. Return target_cell as {sheet, coordinate} for that cell and new_value as the parsed number. (Do not substitute the specific currencies, cell, or number from any prior run \u2014 derive them from this input's slide_texts and cells_dump.)"
    },
    {
      "instruction": "Overwrite target_cell with new_value in the embedded workbook, leaving all formula cells untouched, then recalculate cached formula values with the skill's LibreOffice-backed recalc script. The program refuses if target_cell holds a formula and fails if recalculation reports any Excel errors.",
      "program": "apply_update_and_recalc"
    },
    {
      "instruction": "Repack the unpacked directory into the output .pptx, leaving every other part of the presentation unchanged.",
      "program": "repack"
    }
  ],
  "verification": [
    {
      "instruction": "Open the repacked output .pptx, extract its embedded workbook, and confirm target_cell now equals new_value and that the matrix's formula cells are still stored as formulas (none were converted to hardcoded values). This mirrors the read-back the execution performed on its result.",
      "program": "verify_output"
    }
  ]
}
\end{lstlisting}
    \caption{Opus~4.8 capsule for \texttt{exceltable-in-ppt} (part 2 of 2).}
    \label{fig:example-capsule-multistep-2}
  \end{minipage}
\end{figure*}

\begin{figure*}[p]
  \begin{minipage}[t][\dimexpr\textheight-2pt\relax][t]{\textwidth}
\begin{lstlisting}
#!/usr/bin/env python3
import json, os, re, sys
from pathlib import Path

with open(os.environ['SKILLCACHE_INPUT']) as f:
    inp = json.load(f)
input_pptx = inp['input_pptx']
unpacked_dir = inp['unpacked_dir']
assets = os.environ['SKILLCACHE_ASSETS']

unpack = os.path.join(assets, 'pptx', 'ooxml', 'scripts', 'unpack.py')
if not os.path.exists(unpack):
    sys.exit(f'unpack.py not found at {unpack}')
if not os.path.exists(input_pptx):
    sys.exit(f'input pptx not found: {input_pptx}')

import subprocess
r = subprocess.run([sys.executable, unpack, input_pptx, unpacked_dir],
                   capture_output=True, text=True)
if r.returncode != 0:
    sys.stderr.write(r.stdout + r.stderr)
    sys.exit(f'unpack failed rc={r.returncode}')

root = Path(unpacked_dir)

# Locate the embedded Excel workbook under ppt/embeddings/
embeds = sorted(root.glob('ppt/embeddings/*.xlsx'))
if not embeds:
    sys.exit('no embedded .xlsx found under ppt/embeddings/')
if len(embeds) > 1:
    sys.stderr.write('multiple embedded workbooks found; using the first: '
                     + ', '.join(str(e) for e in embeds) + '\n')
embedded = embeds[0]
embedded_rel = str(embedded.relative_to(root))

# Extract slide text runs (<a:t>...</a:t>) in document order across all slides
slide_texts = []
for slide in sorted(root.glob('ppt/slides/slide*.xml')):
    txt = slide.read_text(encoding='utf-8')
    for m in re.findall(r'<a:t>([^<]*)</a:t>', txt):
        slide_texts.append(m)
if not slide_texts:
    sys.stderr.write('warning: no <a:t> text runs found in slides\n')

# Dump every non-empty cell: formula/value and cached (data_only) value
from openpyxl import load_workbook
wb = load_workbook(embedded)
wbd = load_workbook(embedded, data_only=True)
cells = []
for ws in wb.worksheets:
    wsd = wbd[ws.title]
    for row in ws.iter_rows():
        for c in row:
            if c.value is not None:
                cells.append({
                    'sheet': ws.title,
                    'coordinate': c.coordinate,
                    'formula_or_value': c.value,
                    'cached_value': wsd[c.coordinate].value,
                })

print(json.dumps({
    'status': 'complete',
    'result': {
        'embedded_xlsx_rel': embedded_rel,
        'slide_texts': slide_texts,
        'cells_dump': cells,
    }
}))
\end{lstlisting}
    \caption{Program \texttt{unpack\_and\_inspect} of the capsule in Figure~\ref{fig:example-capsule-multistep}.}
    \label{fig:multistep-unpack}
  \end{minipage}
\end{figure*}

\begin{figure*}[p]
  \begin{minipage}[t][\dimexpr\textheight-2pt\relax][t]{\textwidth}
\begin{lstlisting}
#!/usr/bin/env python3
import json, os, subprocess, sys
from pathlib import Path

with open(os.environ['SKILLCACHE_INPUT']) as f:
    inp = json.load(f)
unpacked_dir = inp['unpacked_dir']
embedded_rel = inp['embedded_xlsx_rel']
target = inp['target_cell']
new_value = inp['new_value']
assets = os.environ['SKILLCACHE_ASSETS']

embedded = str(Path(unpacked_dir) / embedded_rel)
if not os.path.exists(embedded):
    sys.exit(f'embedded workbook not found: {embedded}')

from openpyxl import load_workbook
wb = load_workbook(embedded)
sheet = target['sheet']
coord = target['coordinate']
if sheet not in wb.sheetnames:
    sys.exit(f'sheet {sheet!r} not in workbook sheets {wb.sheetnames}')
ws = wb[sheet]
cur = ws[coord].value
if isinstance(cur, str) and cur.startswith('='):
    sys.exit(f'target cell {sheet}!{coord} holds a formula ({cur!r}); refusing to overwrite a formula cell')
ws[coord] = new_value
wb.save(embedded)

# Recalculate cached formula values with the skill's LibreOffice-backed script
recalc = os.path.join(assets, 'xlsx', 'recalc.py')
if not os.path.exists(recalc):
    sys.exit(f'recalc.py not found at {recalc}')
r = subprocess.run([sys.executable, recalc, embedded], capture_output=True, text=True)
sys.stderr.write(r.stdout)
sys.stderr.write(r.stderr)
if r.returncode != 0:
    sys.exit(f'recalc failed rc={r.returncode}')
try:
    rc = json.loads(r.stdout)
except Exception as e:
    sys.exit(f'could not parse recalc output: {e}')
if rc.get('status') not in ('success',):
    # errors_found or missing -> surface and fail
    sys.exit(f'recalc reported problems: {json.dumps(rc)}')

print(json.dumps({'status': 'complete', 'result': {
    'updated_cell': f'{sheet}!{coord}',
    'written_value': new_value,
    'total_formulas': rc.get('total_formulas'),
    'total_errors': rc.get('total_errors', 0),
}}))
\end{lstlisting}
    \caption{Program \texttt{apply\_update\_and\_recalc} of the capsule in Figure~\ref{fig:example-capsule-multistep}.}
    \label{fig:multistep-apply}
  \end{minipage}
\end{figure*}

\begin{figure*}[p]
  \begin{minipage}[t][\dimexpr\textheight-2pt\relax][t]{\textwidth}
\begin{lstlisting}
#!/usr/bin/env python3
import json, os, subprocess, sys

with open(os.environ['SKILLCACHE_INPUT']) as f:
    inp = json.load(f)
unpacked_dir = inp['unpacked_dir']
output_pptx = inp['output_pptx']
assets = os.environ['SKILLCACHE_ASSETS']

pack = os.path.join(assets, 'pptx', 'ooxml', 'scripts', 'pack.py')
if not os.path.exists(pack):
    sys.exit(f'pack.py not found at {pack}')
if not os.path.isdir(unpacked_dir):
    sys.exit(f'unpacked dir not found: {unpacked_dir}')

r = subprocess.run([sys.executable, pack, unpacked_dir, output_pptx],
                   capture_output=True, text=True)
sys.stderr.write(r.stdout)
sys.stderr.write(r.stderr)
if r.returncode != 0:
    sys.exit(f'pack failed rc={r.returncode}')
if not os.path.exists(output_pptx) or os.path.getsize(output_pptx) == 0:
    sys.exit(f'output pptx missing or empty: {output_pptx}')

print(json.dumps({'status': 'complete', 'result': {
    'output_pptx': output_pptx,
    'size_bytes': os.path.getsize(output_pptx),
}}))
\end{lstlisting}
    \caption{Program \texttt{repack} of the capsule in Figure~\ref{fig:example-capsule-multistep}.}
    \label{fig:multistep-repack}
  \end{minipage}
\end{figure*}

\begin{figure*}[p]
  \begin{minipage}[t][\dimexpr\textheight-2pt\relax][t]{\textwidth}
\begin{lstlisting}
#!/usr/bin/env python3
import json, os, sys, tempfile, zipfile

with open(os.environ['SKILLCACHE_INPUT']) as f:
    inp = json.load(f)
output_pptx = inp['output_pptx']
embedded_rel = inp['embedded_xlsx_rel']
target = inp['target_cell']
new_value = inp['new_value']

if not os.path.exists(output_pptx):
    sys.exit(f'output pptx not found: {output_pptx}')

tmp = tempfile.mkdtemp()
zipfile.ZipFile(output_pptx).extractall(tmp)
embedded = os.path.join(tmp, embedded_rel)
if not os.path.exists(embedded):
    sys.exit(f'embedded workbook not present in output: {embedded_rel}')

from openpyxl import load_workbook
wb = load_workbook(embedded)
wbd = load_workbook(embedded, data_only=True)
ws = wb[target['sheet']]
coord = target['coordinate']
val = ws[coord].value
if val != new_value:
    sys.exit(f'target cell {coord} is {val!r}, expected {new_value!r}')

# Confirm formula cells remain formulas (none were hardcoded to values)
formulas = []
for row in ws.iter_rows():
    for c in row:
        if isinstance(c.value, str) and c.value.startswith('='):
            formulas.append(f"{target['sheet']}!{c.coordinate}")

print(json.dumps({'status': 'complete', 'result': {
    'target_cell_value': val,
    'formula_cells_preserved': formulas,
    'formula_cell_count': len(formulas),
}}))
\end{lstlisting}
    \caption{Program \texttt{verify\_output} of the capsule in Figure~\ref{fig:example-capsule-multistep}.}
    \label{fig:multistep-verify}
  \end{minipage}
\end{figure*}